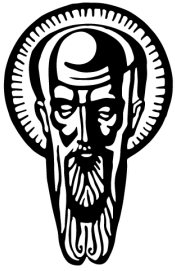

Sofia University "St. Kliment Ohridski"

Faculty of Mathematics and Informatics

# Voice Signal Processing for Machine Learning. The Case of Speaker Isolation

Overview and Evaluation of Decomposition Methods
Applied to the Input Signal of Voice Processing ML Models.
The Use Case of the Speaker Isolation Problem

Radan Ganchev
MSc. Informatics

Thesis advisor: Kalin Georgiev
Reviewer: Trifon Trifonov

Sofia, 2021

# Abstract

The widespread use of automated voice assistants along with other recent technological developments have increased the demand for applications that process audio signals and human voice in particular. Voice recognition tasks are typically performed using artificial intelligence and machine learning models. Even though end-to-end models exist, properly pre-processing the signal can greatly reduce the complexity of the task and allow it to be solved with a simpler ML model and fewer computational resources. However, ML engineers who work on such tasks might not have a background in signal processing which is an entirely different area of expertise.

The objective of this work is to provide a concise comparative analysis of Fourier and Wavelet transforms that are most commonly used as signal decomposition methods for audio processing tasks. Metrics for evaluating speech intelligibility are also discussed, namely Scale-Invariant Signal-to-Distortion Ratio (SI-SDR), Perceptual Evaluation of Speech Quality (PESQ), and Short-Time Objective Intelligibility (STOI). The level of detail in the exposition is meant to be sufficient for an ML engineer to make informed decisions when choosing, fine-tuning, and evaluating a decomposition method for a specific ML model. The exposition contains mathematical definitions of the relevant concepts accompanied with intuitive non-mathematical explanations in order to make the text more accessible to engineers without deep expertise in signal processing. Formal mathematical definitions and proofs of theorems are intentionally omitted in order to keep the text concise.

# Table of Contents

# 1. Introduction

Throughout history there have been many applications of signal processing techniques in a wide variety of areas – from physics and astrology, through medicine and biology, to finances, art, and entertainment. The groundbreaking work of Joseph Fourier from the beginning of the 19th century laid the foundations of modern signal analysis. Fourier showed that any signal can be decomposed as a sum of its frequency components in the form of sine and cosine functions. Since then, the duality of the time and frequency representations of a signal has been explored in various ways, both analytically and experimentally. This has led to the development of wavelet theory a century later – a notable addition to the family of time-to-frequency domain transformations.

The technological advancements of the 21st century are once again putting the theory of signal analysis to the test. An ever-increasing amount of devices generate various new types of signals that need to be analyzed and processed, often in real-time. Personal health trackers, voice assistants, hearing aids, and medical equipment are just a few examples. In addition, the rise of artificial intelligence (AI) and machine learning (ML) models is allowing engineers to tackle new types of complicated problems which were unthinkable in the near past. Asking Siri to turn on the lamps or to tell you which song is currently playing on the radio is no longer the matter of science fiction but a daily routine. The power and simplicity of human voice as a medium for human-machine interaction is motivating various new research efforts in the field of voice processing. These efforts often rely on signal processing techniques working hand-in-hand with statistical inference methods.

Voice recordings are a special type of audio signals with some additional characteristics. Firstly, they are non-periodic. Therefore, the standard Fourier Transform (FT) which provides the frequency decomposition of the whole signal, is not very useful in most cases. A windowed version needs to be used instead – Short-Time Fourier Transform (STFT), which can not only identify frequencies but also localize them in time. Applying a Wavelet Transform (WT) is another alternative, since it is localized in time by definition.

Another characteristic of the human voice is its frequency range. Fundamental voice frequencies range from 100-120 Hz for males to 200-240 Hz for females to around 300 Hz for children. These frequencies, however, apply only to vowels. Consonants are created by air blockages and noise sounds formed by the passage of air through the throat and mouth. Their frequencies lie above 500 Hz and can reach up to 4 kHz. Since consonants contribute more to speech intelligibility than vowels, the range between 2 and 4 kHz is considered the most important when interpreting speech [1]. This is also something to consider when choosing or fine-tuning a frequency decomposition method for a particular voice processing task.

These intricacies of voice signals show that choosing the most suitable mathematical transformation to apply prior to statistical inference is not always trivial. ML engineers often resort to using STFT as the most common and well-known option. But even though it is powerful and versatile, STFT also has hyperparameters which, if not fine-tuned for the task, may lead to poor performance. Wavelet Transform (WT) remains seldomly used in ML models and Wavelet Packet Transform (WPT) even more so. The reason for this might be the fact that understanding these transformations requires a solid mathematical background. Some good textbooks on the topic exist but most of them contain an amount of mathematical details which may seem overwhelming to an ML engineer with a specialization and interests in an entirely different subfield of mathematics.

It is worth mentioning that end-to-end ML models are also used and have become remarkably accurate for some specific voice-processing tasks such as clean speech recognition. Such models work with the audio signal directly, without the prior application of a mathematical transformation or extraction of features from it. In many cases, however, these models contain millions of trainable parameters and require enormous amounts of labeled data and computing power to train. Decomposing the signal and analyzing its frequency components, on the other hand, can make some features of the signal more explicit and easy to detect. This in turn can lead to ML models of smaller complexity and size. While this approach reduces the amount of resources necessary for training the model, it also increases the difficulty and prior knowledge required for building it.

## 1.1. Goal

The goal of the current work is to provide a concise comparative analysis of modern signal decomposition methods and metrics for evaluating speech intelligibility, commonly used in voice processing ML models. The level of detail in the exposition is meant to be sufficient for an ML engineer to make informed decisions when choosing and fine-tuning a decomposition method for a specific ML model. Formal mathematical definitions and proofs of theorems are intentionally omitted in order to keep the text concise.

The decomposition methods are analyzed in the context of an implementation of an experiment illustrating how they can be compared and fine-tuned for a specific task.

## 1.2. Focus and Scope

This thesis focuses mainly on the mathematical methods and transformations which are applied prior to statistical inference, as well as on metrics typically used when building voice processing ML models. The structure and the variety of the available relevant ML models is outside the current scope.

Practical examples throughout the exposition use the Python programming language since currently it is the most commonly used language for building ML models.

It is important to note that while the methods and tools described here are treated in the context of voice processing problems, their application is not limited to such problems. On the contrary, most tools can be used in a wide variety of tasks. STFT, WT and WPT in particular, can be applied to any signal processing task, even for signals of higher dimensions, such as images.

## 1.3. Thesis Structure

The exposition starts with a short overview of mathematical and signal processing terms and notations used in later sections. The main part of the thesis follows – a description of the commonly used signal decomposition methods and metrics used when solving voice processing tasks. The work concludes with a case study where the specific problem of speaker isolation is treated in some detail in order to demonstrate the practical applications of the previous theoretical sections.

> Throughout the work, the author's notes are outlined in sections on a darker background. They are prefixed with one of the following keywords:
> - ***Highlight:*** These sections contain summaries or facts which the author considers to be key pieces of knowledge regarding the practical applications of the corresponding matter.
> - ***Intuition:*** These sections contain the author's intuitive understanding of the matter which may help the reader build a better intuition of their own.

# 2. Theoretical Background and Notations

The mathematical details in the exposition are kept to a minimum, but the reader is still expected to have some basic mathematical background and feel comfortable with function notations, summations, integrals, matrix and vector multiplications. This section contains a short description of some of the key theoretical terms and concepts used throughout this work.

## 2.1. Continuous Signals

When referring to continuous audio signals, we typically mean continuous real-valued functions in the time domain $f : \mathbb{R} \to \mathbb{R}$. They are expected to have "good" mathematical properties. Intuitively speaking, this means that they should be "smooth" and not have discontinuities or diverge to infinity. Fourier and Wavelet analyses require the signals to have finite energy, i.e., to be square-integrable:

$$\|f\|^2 = \int |f(t)|^2 dt < \infty,$$

where $\|f\| = \sqrt{\langle f, f \rangle}$ is the norm of $f$ in a Hilbert space with inner product:

$$\langle f, g \rangle = \int f(t)\overline{g(t)} dt$$

(The norm and inner product definitions are intentionally given in a form which can also be applied to complex-valued functions. In the real-valued case, if $g : \mathbb{R} \to \mathbb{R}$, then its complex conjugate is simply the function itself: $\overline{g(t)} = g(t)$.)

The set of square-integrable real-valued functions is traditionally denoted as $L^2(\mathbb{R})$ and the norm $\|f\|$ is also known as $L^2$-norm. The set of square-integrable functions defined on a finite interval $[a, b] \subset \mathbb{R}$ is denoted $L^2[a, b]$.

## 2.2. Bases in Function Spaces

If $F$ is an inner product function space (such as $L^2(\mathbb{R})$ or $L^2[0, 1]$, for example), then a basis of $F$ is any subset $B \subset F$ for which every $f \in F$ can be uniquely expressed as a linear combination of elements of $B$. If the elements of $B$ are orthogonal, i.e. $\langle a, b \rangle = 0$ for $a, b \in B, a \neq b$, then $B$ is called an *orthogonal basis*. If additionally all basis functions have unit norm $\|a\| = 1, \forall a \in B$, then $B$ is an *orthonormal basis*.

> ***Highlight:*** Mathematically, decomposing a signal often consists of finding its coordinates in a given basis. Thus the search for the most optimal decomposition method typically resorts to finding the most optimal basis in which to decompose the signal.

Orthonormal bases have "good" mathematical properties which are often desired, but requiring orthonormality can be limiting with regard to the methods used for constructing the basis itself. Therefore, in some cases the orthogonality requirement is relaxed and biorthogonal bases are constructed instead.

A *biorthogonal system* is a pair of indexed families $A = \{a_n\}_{n \in \mathbb{Z}} \subset F$ and $B = \{b_n\}_{n \in \mathbb{Z}} \subset F$ such that $\langle a_i, b_i \rangle = 1$ and $\langle a_i, b_j \rangle = 0, i \neq j$. If $A$ and $B$ are (possibly non-orthogonal) bases of $F$, then the biorthogonal system is called a *biorthogonal basis* of $F$.

> ***Intuition:*** In a biorthogonal basis it is obvious that if $A \equiv B$, then $A$ is an orthonormal basis of $F$. Thus, building a signal decomposition method as a biorthogonal system can be viewed as "splitting" the orthonormality requirement across two different sets of elements which together act as an orthonormal basis.

## 2.3. Frequency Domain

When a signal is converted from time domain to frequency domain, it is represented as a weighted sum of frequencies. The frequencies themselves are periodic functions in time, typically sines and cosines. Periodic functions have two important characteristics – magnitude and phase.

In many cases when audio signals are analyzed, the magnitude carries most of the relevant information and the phase information is not analyzed, but only stored for the purpose of reconstructing the signal if necessary.

> ***Intuition:*** The magnitude represents "how loud" the frequency is and the phase

> describes how it is shifted in time. Since frequencies are infinitely defined periodic functions, shifting them in time does not change their nature. Thus the phase is often discarded as irrelevant unless it is important for a specific task to align the different frequencies precisely.

Since in the frequency domain the signal is described by two independent characteristics at any given point in time, this information cannot be conveniently represented by a single real number. Thus, conversion from time to frequency domain typically involves switching from real-valued to complex-valued functions. If a complex-valued periodic function is expressed in polar coordinate form, then the absolute value of the radial coordinate would denote the magnitude of the frequency and the angular coordinate would denote the phase. Therefore, in mechanical engineering frequencies are typically represented by complex exponentials of the form $e^{i\omega t}$ where $i$ is the imaginary unit, $\omega$ is the (angular) frequency, and $t$ is the time parameter. We can use Euler's formula to convert this representation to its trigonometric equivalent which makes its periodic nature more obvious:

$$e^{i\omega t} = \cos \omega t + i \sin \omega t$$

Angular frequencies are often used instead of linear frequencies in the field of physics and sometimes in signal processing. The letters used to denote these two types of frequencies are typically $\omega$ (lowercase omega) and $f$, respectively. The relationship between them is $\omega = 2\pi f$. So if one prefers to use linear frequencies, the above equation can also be written in the form:

$$e^{2\pi i f t} = \cos 2\pi f t + i \sin 2\pi f t$$

Throughout this work we will use the former notation in formulas since it is more concise. However, in some cases it makes more sense to discuss linear frequencies since they are more intuitive in practice. In such cases we will explicitly refer to the frequency as "linear" to avoid confusion.

## 2.4. Discrete Signals

Continuous signals are typically used in theoretical treatises on the topic, but for practical purposes we need to switch to a discrete and even finite domain. This is necessary because signals in computer memory are typically represented by a finite sequence of samples and are discrete by definition. Fortunately, most theoretical results in continuous signal analysis have also been transferred to the discrete domain. Proving that a result in continuous time is also valid in discrete time may not be trivial. However, if the result has already been proven and if we are concerned only with its practical aspects, switching to discrete time is often as easy as replacing continuous with discrete functions, integrals with summations, etc.

We will denote discrete functions with bold letters, e.g. $\boldsymbol{h} : \mathbb{Z} \to \mathbb{R}$, and their parameter will be specified in square brackets to make it more obvious that it is an integer: $\boldsymbol{h}[t], t \in \mathbb{Z}$. The discrete $L^2$ norm is thus given by:

$$\|\boldsymbol{h}\|^2 = \sum_t \boldsymbol{h}[t]^2$$

If $\boldsymbol{h}[t] = 0$ for $t \notin [0, L-1], L \in \mathbb{N}$ then $\boldsymbol{h}$ can be represented simply as a vector $h \in \mathbb{R}^L$. When switching to a finite domain some length-related conditions typically appear. For example, when applying Discrete Wavelet Transform (DWT), the length of the signal is required to be a power of two. In practice this is achieved by padding the signal, e.g. with zeros or by repetition. Such technicalities are specific to certain algorithms and are outside the scope of this work. When we treat discrete signals throughout this work, we will assume that all requirements on their length are implicitly satisfied. In practice, available programming libraries typically handle padding internally and expose configuration options with which the programmer can specify what type of padding should be used.

### 2.4.1. Frequency

The definition of "frequency" in the discrete domain is slightly different than in the continuous domain. For a continuous periodic signal we typically define its linear frequency to be the number of oscillations *per second*. This is a theoretically unlimited positive quantity. In the discrete case, however, the number of samples per second may vary and it is typically treated as an external parameter. Thus the frequency is defined as the number of oscillations *per time step (i.e., per sample)*. So the maximum frequency is achieved by the signal which alternates signs at every sample $\boldsymbol{x}[t] = (-1)^t$ and its linear frequency is equal to $1/2$. The corresponding maximal angular frequency is thus $\omega = \pi$. This maximum frequency can also be represented by the complex exponential $e^{i\pi t}, t \in \mathbb{Z}$.

> ***Highlight:*** Frequencies of discrete signals are finite and range from $0$ to $\pi$ (if viewed as angular frequency).

To map the discrete frequency of a signal to its actual frequency, we need to know the sampling rate. For example, if the sampling rate of the signal is 8 kHz and its discrete linear frequency is $1/2$, then its actual frequency is 4 kHz because it contains 4000 oscillations per second. Inversely, if we have a continuous signal of frequency $NHz$, then we would need a sampling rate of *at least* $2NHz$ in order to capture it without loss of information. This is known as the *Nyquist–Shannon sampling theorem*. The minimal sampling rate of $2NHz$ necessary for representing a frequency of $NHz$ is sometimes called the *Nyquist rate*. Also, the maximum frequency of $NHz$ which can be represented by a sampling rate of $2NHz$ is called the *Nyquist frequency* for this sampling rate.

## 2.5. Invertibility

In most cases when we apply some transformation on a signal, we would like it to be invertible, even if we do not intend to reconstruct the signal later. This is because invertible transformations are lossless by definition, so they preserve all of the signal's information.

Say we have a signal $f : \mathbb{R} \to \mathbb{R}$ and a transformation $T$ which transforms $f$ to some other representation $\hat{f}$:

$$T\{f\} = \hat{f}$$

We say that $T$ is invertible if there exists an inverse transformation $T^{-1}$ such that:

$$T^{-1}\{\hat{f}\} = f$$

Sometimes we allow transformations to introduce a time delay. So, for example, we may relax the above requirement to:

$$T^{-1}\{\hat{f}\}(t) = f(t - \tau)$$

For some constant $\tau \in \mathbb{R}$.

Invertibility can be described analogously for transformations on discrete signals via a trivial substitution of $f$ above with a discrete function.

> **Example:** Let $B = \{b_i\}_{i \in \mathbb{Z}} \subset L^2(\mathbb{R})$ be a basis of $L^2(\mathbb{R})$ and let $T\{f\} = \{a_i\}_{i \in \mathbb{Z}}$ be a transformation which maps every function $f \in L^2(\mathbb{R})$ to its coordinates $\{a_i\}$ in the basis $B$. Then $T$ is invertible and its inverse is:
>
> $$T^{-1}\{\{a_i\}_{i \in \mathbb{Z}}\} = \sum_{i \in \mathbb{Z}} a_i b_i = f$$
>
> Thus, all basis decompositions can be viewed as invertible transformations.

## 2.6. Convolution

Convolution is a mathematical operation[1] with which one function (or vector in the discrete case) acts upon another. It is typically denoted by the asterisk symbol $*$. If $f, g : \mathbb{R} \to \mathbb{R}$, in its continuous form, it is defined as follows:

$$f(t) * g(t) = (f * g)(t) = \int f(\tau) g(t - \tau) d\tau$$

In the discrete case, if $\boldsymbol{x}, \boldsymbol{h} : \mathbb{Z} \to \mathbb{R}$, convolution is defined as:

$$\boldsymbol{x}[t] * \boldsymbol{h}[t] = (\boldsymbol{x} * \boldsymbol{h})[t] = \sum_n \boldsymbol{x}[n] \boldsymbol{h}[t - n]$$

If one of the operands, say $g$ or $\boldsymbol{h}$ above, describes a signal-processing filter for example, then the application of the convolution operation can be viewed as the transformation which this filter applies to the incoming signal $f$ or $\boldsymbol{x}$, respectively. In the discrete case, if $\boldsymbol{h}$ has finite support, then it is trivial to implement a software or hardware device which can perform the convolution of $\boldsymbol{h}$ and any incoming signal $\boldsymbol{x}$. This is the reason why convolution plays a central role in signal processing.

---

[1] Mathematical convolution is similar but slightly different from the convolution operator used in machine learning models such as Convolutional Neural Networks (CNNs), for example. When we speak of convolution throughout this work, we will refer to mathematical convolution.

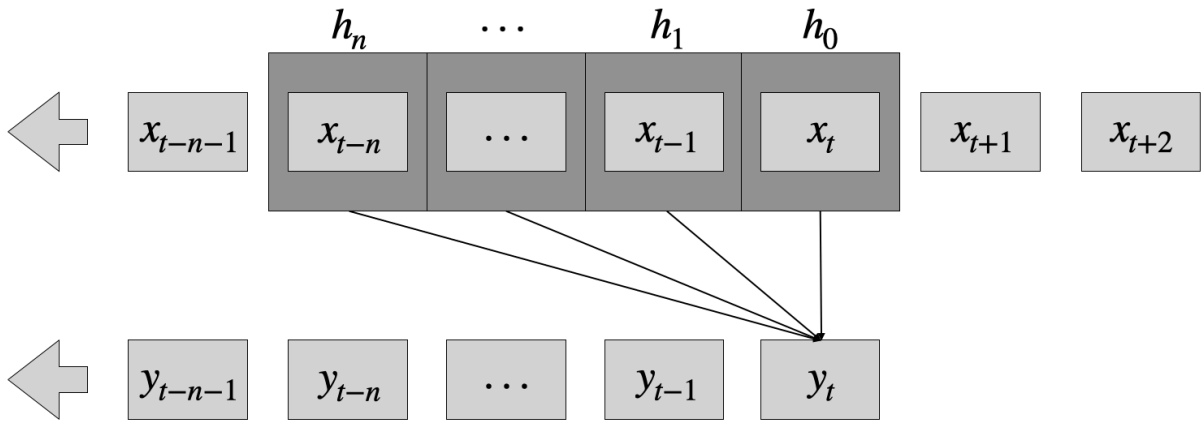

*Figure 1. A filter described by the vector $h \in \mathbb{R}^n$ acts upon incoming signal $x$ by convolution and produces the output signal $y$.*

> ***Highlight:*** Convolution and filters are tightly related. If a signal transformation can be described as a combination of one or more convolutions with finite vectors, then it can be implemented as a series of software and/or hardware filters with relatively low complexity and high efficiency.

# 3. Signal Decompositions

Short-Time Fourier Transform (STFT) is the most widely used decomposition method in recent developments in the area of voice processing. Wavelet Transform (WT) and Wavelet Packet Transform (WPT) have also been used in speech processing although they are typically applied specifically to solving problems such as de-noising [2] or compression [3]. We will provide a short overview of these three decomposition methods in the following sections and examine some of their properties relevant to their application prior to statistical inference in voice processing problems.

## 3.1. Short-Time Fourier Transform

Fourier analysis aims to provide a different view over a given function which in some cases may be easier to interpret or analyze than the function itself. A speech signal, for example, is typically represented by a continuous real function in time, oscillating in a non-periodic fashion. By applying a Fourer transform to it, we can convert it to a "weighted sum" of frequencies represented by a basis of wave functions of the form $e^{i\omega t}$ for various angular frequencies $\omega$. In this form we can reason about the audio content of the signal and associate some frequencies with noise and others with human voice, for example. In the context of an ML model such frequency-based dependencies would be easier to learn than working with the original signal in the time domain.

The continuous Fourier transform of a function $f(t) \in L^2(\mathbb{R})$ is given by the formula:

$$\hat{f}(\omega) = \langle f, e^{i\omega t} \rangle = \int f(t) e^{-i\omega t} \, dt$$

The original function $f$ can be reconstructed by applying the inverse transform:

$$f(t) = \frac{1}{2\pi} \int \hat{f}(\omega) e^{i\omega t} \, d\omega$$

When $f(t)$ is defined only on an interval, say $[0, 1]$, then the Fourier transform becomes a decomposition in a Fourier orthonormal basis $\{e^{i 2\pi m t}\}_{m \in \mathbb{Z}}$ of $L^2[0, 1]$ [4]. Each basis function $e^{i 2\pi m t}$ represents the integer linear frequency $m$.

During this transformation, however, we lose information about the *time* at which a given frequency occurred. This is due to the uncertainty principle (the Gabor limit in particular) which states that we cannot simultaneously sharply localize a signal in both the time domain and frequency domain [5]. To mitigate this problem, we can modulate the signal with a window function, essentially breaking it into parts, and then apply Fourier transform to each part separately. This method is known as the "windowed Fourier transform" or "short-time Fourier transform".

For practical applications it makes sense to focus only on the discrete-time STFT. The STFT of a discrete signal $x[t]: \mathbb{Z} \to \mathbb{R}$ is given by the formula:

$$STFT\{x[t]\}(m, \omega) = \sum_{n=-\infty}^{\infty} x[n]w[n-m]e^{-i\omega n}$$

where $w[t]$ is a window function which, by definition, is non-zero only inside a predefined interval (described in detail in the following section). This definition of STFT uses an infinite version of the input signal $x[t]$ and a continuous frequency parameter $\omega$ so it is still not in a practical form for computation purposes. With a few transformations STFT can be converted to a series of Discrete Fourier Transforms (DFT) which, in turn, can be computed via the Fast Fourier Transform (FFT) algorithm. Another approach would be to represent the STFT via a convolution operation and use filter banks for its computation. The exact computation method is not relevant to the current work, so these methods will not be described in detail here.

### 3.1.1. Window functions

As mentioned above, window functions are functions which are non-zero only inside a predefined interval, known as *support* interval. When choosing a window function for STFT we need to consider three main characteristics: shape, support interval, and overlap.

Applying a window to a signal prior to its frequency decomposition causes an undesired side effect known as *spectral leakage*. Spectral leakage is a phenomenon where the power of some frequency "leaks" into neighboring frequencies. Thus the windowed Fourier transform may detect frequencies which are not present in the original signal and reduce the power of actual frequencies which are present in the signal.

The shape of a window function in its support interval determines its spectral leakage properties. Some windows shift most of the leaked power to "nearby" frequencies while others distribute the leaked power across a wider range of frequencies.

The most basic window functions are rectangular windows which are equal to 1 in the whole given interval. Rectangular windows add the smallest amount of total noise due to spectral leakage compared to other window functions. However, the power leaked from rectangular windows spreads widely, to frequencies far from the original one (Figure 2). This makes them a poor choice for voice processing applications.

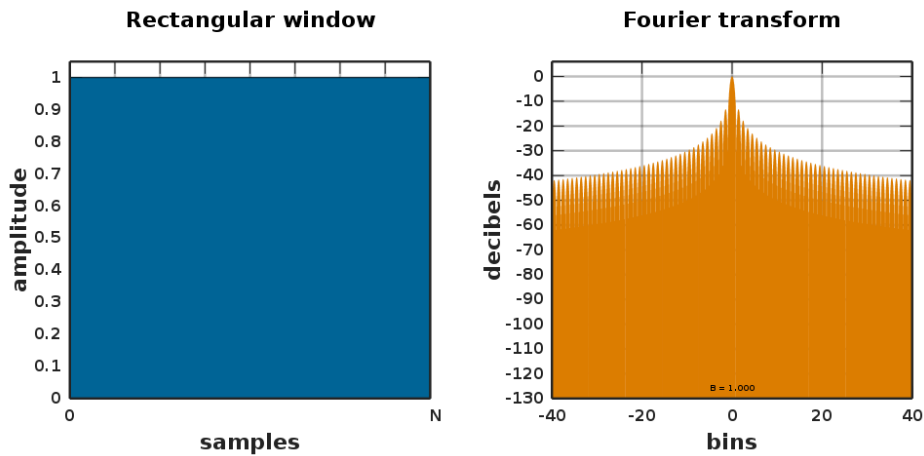

*Figure 2. (Source: Wikipedia) Rectangular window. The Fourier transform on the right shows the window's spectral leakage properties. The main lobe around the central frequency is narrow and the leakage spreads uniformly to the sides.*

Using a window function that diminishes the signal near both ends of the frame helps smoothen the windowed signal and contain the leakage to a narrower area. For example, below is the Hann window named after meteorologist Julius von Hann[2]:

$$\boldsymbol{w}[n] = \begin{cases} 0.5 - 0.5\cos(\frac{2\pi n}{N}), & \text{if } 0 \leq n \leq N \\ 0, & \text{otherwise} \end{cases}$$

---

[2] The Hann window is often erroneously called "hanning window". According to Wikipedia, the confusion was caused by the similarly named Hamming window, proposed by Richard Hamming. The Hamming window has a very similar shape to the Hann window, but it does not reach zero at both ends.

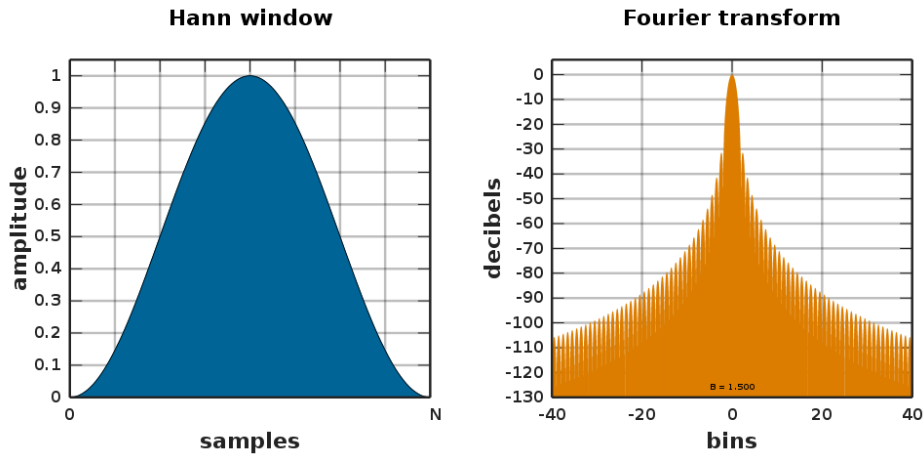

*Figure 3. (Source: Wikipedia) Hann window. The Fourier transform on the right shows the window's spectral leakage properties. The main lobe around the central frequency is relatively wide, but the leakage fades quickly and does not spread to frequencies that are too far from the central one.*

It can be seen from the formula and Figure 3 that Hann windows have a symmetric bell-like shape, reaching 1 in the middle and diminishing to zero near both ends. The simplicity and the smoothing properties of Hann windows make them the default choice for many signal-processing applications, even though other good window functions exist.

Once we have selected the shape of the window function, we need to consider what the window size (interval) and overlap will be. In the discrete case the interval and overlap are often measured in number of samples but sometimes it is more convenient to convert this number to a time interval (usually in milliseconds). This conversion is possible because samples are evenly distributed at a fixed sample rate per second.

Choosing the "correct" interval for the window function is not straightforward. Narrower windows provide better resolution in the time domain (*when* a frequency appeared) but a poorer resolution in the frequency domain (*what* frequency appeared). Using a wider window results in the opposite. Because of this tradeoff, the window size is often chosen empirically based on the specific task. For speech analysis, in [6] it is suggested that a size between 15 and 35 ms is most efficient. However, other texts on the topic, such as [7], suggest window sizes anywhere between 10 and 120 ms.

Overlapping the adjacent time frames is necessary so that we do not lose data during reconstruction. For example, if we use a Hann window with no overlap, the values at the borders between each two adjacent frames will always be zero, because the window function itself is zero at both ends. Furthermore, if there is only a small overlap, say 5% of the window size, the maximum magnitude of the combined signal in the overlapped area between adjacent frames will be reduced. This is because the value of the Hann window near its edges is low, so even after we combine the magnitudes from two adjacent frames, they will still be strictly less than the magnitude of the original signal. Therefore, overlaps of 25% or 50%, or more are common when using Hann windows. For example, a frequently used configuration for voice processing tasks is a 32 ms Hann window with a 8 or 16 ms overlap [8–10].

> *Highlight:* Larger window size means better frequency resolution. Smaller window size means better time resolution. The appropriate size must be chosen depending on the task at hand. For voice processing tasks, a 32 ms Hann window with 8 or 16 ms overlap is a good default.

### 3.1.2. Applications

Many popular Python packages such as SciPy and Librosa offer a function `stft`, which takes the input signal as a parameter, as well as some windowing and padding parameters, and computes the STFT. The output is typically a two-dimensional array containing one row for each analyzed frequency and one column for each time step. The elements of the array are the complex-valued coefficients corresponding to each frequency at each time step.

#### 3.1.2.1. Window Properties

Libraries typically express the STFT window size in number of samples and the window shape as an enumeration of predefined window names. In most cases a Hann window is used as a default.

Care must be taken to verify in the documentation what exactly each parameter represents, because there are subtle differences between the ways in which the window properties are described in different libraries. For example, SciPy uses an `noverlap` parameter to specify the number of overlapping samples for adjacent windows. Librosa, on the other hand, uses a `hop_length` parameter which specifies the distance (again in number of samples) between the starts of adjacent windows. These are two different ways to describe the same underlying characteristic.

### 3.1.2.2. Frequency Resolution

Libraries often use a fast Fourier transform (FFT) algorithm to compute STFT. FFT works with a fixed number of samples at a time, known as the FFT length or FFT size. This size, together with the sampling rate of the signal, determines the frequency resolution of the output, i.e., the number of different frequencies the output contains and how they are spaced apart. If the FFT size is $N$ and the sampling rate is $R$, then the frequency resolution of the STFT will be $R/N$. This means that the frequencies included in the output will be:

$$\{k\frac{R}{N}, k = 0, 1, 2, ..., \lfloor\frac{N}{2}\rfloor\}$$

For example, at sampling rate $4kHz$ and FFT size $128$ samples (equivalent to $32ms$), the frequencies included in the STFT output will be $\{0, 31.25, 62.5, 93.75, ..., 2000\} Hz$ or a total of 65 different frequencies.

Note that the FFT size is a parameter different from the window size and some libraries allow setting them both independently. In most cases they will be equal to each other by default but the window size can be set to a value smaller than the FFT size. In such cases, the FFT will be applied on windowed segments of the signal, but since these segments will be smaller than the FFT size, they need to be padded in some way, for example with zeros.

### 3.1.2.3. Example

Below is a Python example of applying STFT to a sample WAV file using SciPy:

*Listing 1. Example usage of STFT with SciPy. Complete implementation can be found at: https://github.com/rganchev/speech-signal-processing-for-ml*

```
import scipy.io.wavfile
import scipy.signal
import math

def read_wav_file():
  file_path = get_file_path() # obtain a path to a sample WAV file
  return scipy.io.wavfile.read(file_path)

fs, wav = read_wav_file()
```

```
print('Sampling frequency: %i Hz' % fs)
print('Number of samples: %i' % wav.shape[0])

win_size = math.floor(0.032 * fs) # use a 32ms window size
f, t, coeffs = scipy.signal.stft(wav, fs=fs, window='hann',
                                 nperseg=win_size,
                                 noverlap=win_size / 2)
print('Analyzed frequencies (Hz): %s' % f)
print('Time steps (seconds): %s' % t)
print('Output dimensions: %s' % (coeffs.shape,))
```

Output (may vary depending on the used WAV file):

```
Sampling frequency: 16000 Hz
Number of samples: 959669
Analyzed frequencies (Hz): [0.    31.25   62.5  ...  7937.5  7968.75  8000.]
Time steps (seconds): [0.    0.016   0.032  ...  59.952  59.968  59.984]
Output dimensions: (257, 3750)
```

### 3.1.2.4. Spectrograms

STFT produces a two-dimensional array of coefficients representing the frequency spectrum of the signal at each time step. This spectrum is typically visualized via plots called *spectrograms*. Spectrograms are simply heat maps of the STFT coefficients. Time steps (time windows) are arranged along the *x* axis and frequencies – along the *y* axis. Each point $(x, y)$ on the spectrogram represents the magnitude of the coefficient of frequency $y$ at time step $x$. The larger the magnitude is, the brighter its representation on the heatmap, which represents a stronger correlation of the signal content at this time step with the corresponding frequency. An example of such a spectrogram is displayed in Figure 4. The source code used to generate this spectrogram is given in the following code listing.

*Listing 2. Plotting a spectrogram. Complete implementation can be found at:*
*https://github.com/rganchev/speech-signal-processing-for-ml*

```
import matplotlib.pyplot as plt
import math

def plot_waveform(ax, wav):
  ax.plot(wav)
```

```
    ax.set(title='Waveform', xlabel='samples', ylabel='amplitude')

def plot_spectrogram(ax, wav, fs):
    win_size = math.floor(0.032 * fs) # use a 32ms window size
    ax.specgram(wav, Fs=fs, NFFT=win_size, noverlap=win_size / 2)
    ax.set(title='Spectrogram', xlabel='time', ylabel='frequency')

fs, wav = read_wav_file() # same as in Listing 1.
wav = wav[:65536] # Analyze only the first few seconds of the wav file

_, (ax1, ax2) = plt.subplots(nrows=2)
plot_waveform(ax1, wav)
plot_spectrogram(ax2, wav, fs)
```

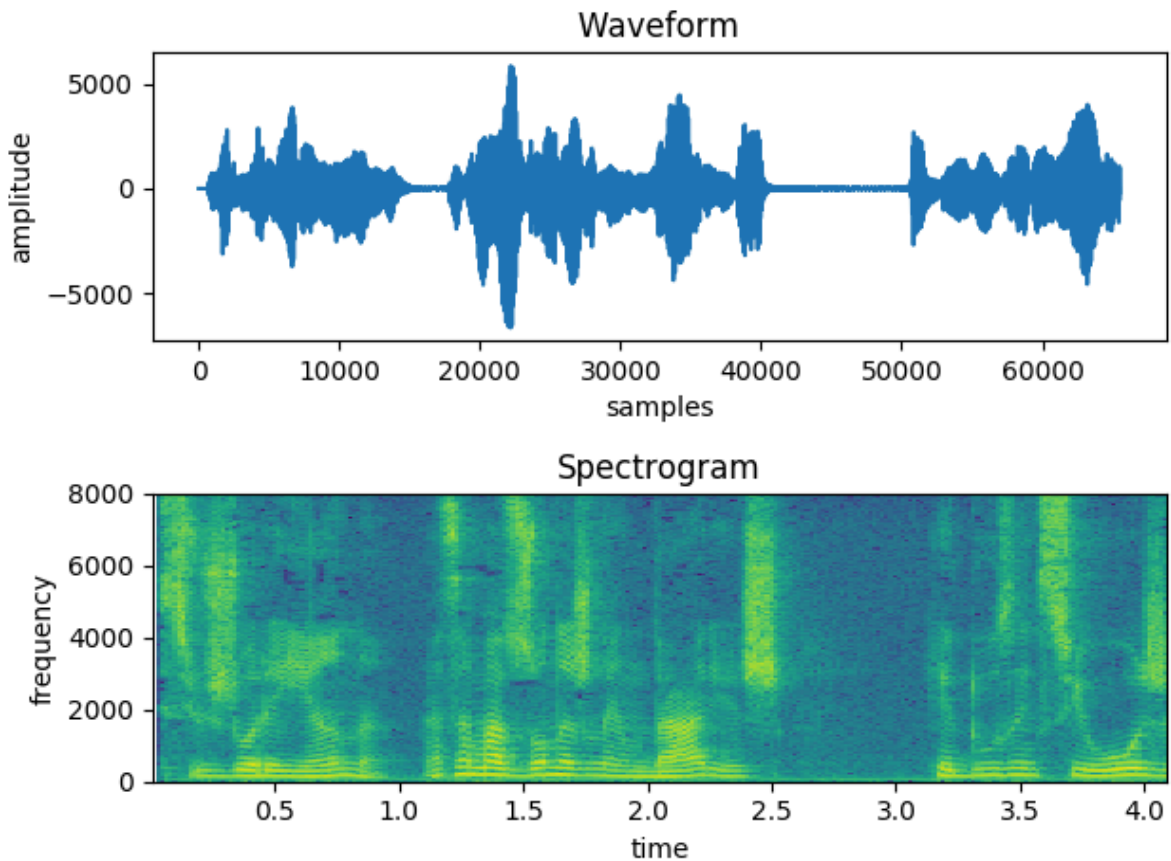

*Figure 4. A voice recording displayed as a raw waveform at the top along with its corresponding spectrogram at the bottom. The recording contains 65,536 samples sampled at 16kHz (i.e. a total length of about 4 seconds). The spectrogram was generated using a Hann window of length 32ms (512 samples) with 16ms (256 samples) overlap.*

By observing the waveform and the spectrogram in Figure 4 the following features may be deduced:
- The maximum displayed frequency is 8 kHz which is the Nyquist frequency for a 16 kHz sampling rate.
- The waveform contains mostly silence between the 40,000th and 50,000th sample. As expected, the corresponding area in the spectrogram is mostly dark.
- The highest-energy frequencies (brightest pixels) lie near the bottom of the spectrum where we expect the base human voice frequencies to be - between 100 and 300 Hz.
- If one listens to the voice recording and pays attention to the location of consonants, it will become obvious that they correspond to the lighter regions of higher frequencies in the spectrogram. This fact is most apparent around the 2.5 second mark where the recording contains the sound "sh" [ʃː] and the spectrogram contains mostly frequencies above 2 kHz respectively.

> ***Highlight:*** Spectrograms are useful for manually analyzing the content of signals, for example during the data exploration phase of building an ML model. In Python, the most widely used plotting library Matplotlib contains a `specgram` method which performs STFT internally and plots the spectrogram directly from the raw audio content.

## 3.2. Wavelet Transform

The Fourier transform can be viewed as a decomposition of a continuous function into a basis of sinusoidal wave functions with infinite support. Wave*lets,* on the other hand, are wave functions with finite support, i.e., they are zero outside of a finite interval. Broadly speaking, if we choose a wavelet (so called "mother" wavelet), we can form a basis of $L^2(\mathbb{R})$ by taking all of its translations and dilations. An example of the dilations of one such wavelet is shown in Figure 5. Translations of the dilated wavelet are formed by "sliding" each dilation along the x-axis.

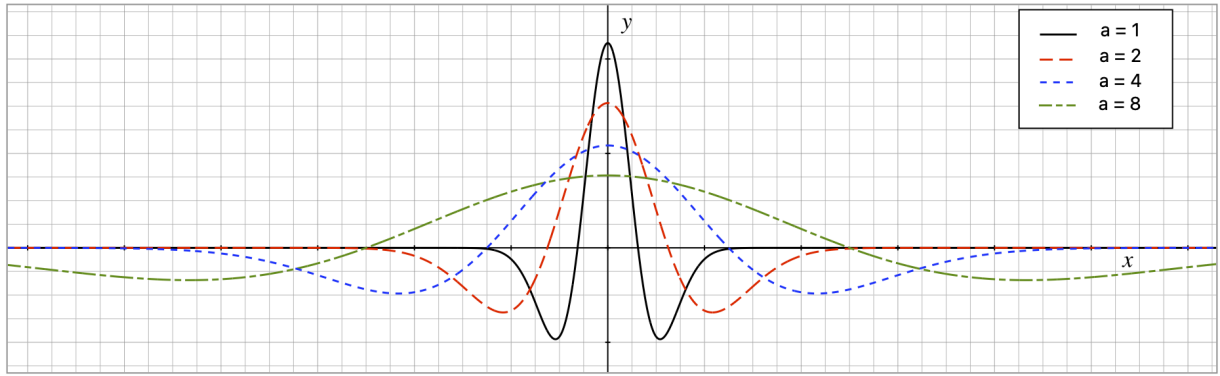

*Figure 5. Dilations of the Ricker wavelet (a.k.a. Mexican hat wavelet). The mother wavelet (scaling factor a = 1) is shown with a solid black line. The rest of the lines show dilations at various scales.*

Depending on the choice of mother wavelet, the basis may have some desirable properties such as orthogonality or orthonormality. The decomposition of a function in such a wavelet basis is called Wavelet Transform (WT) or Wavelet Decomposition. There are several theoretical advantages of wavelet decomposition compared to Fourier transform:
- Wavelets have finite support so WT has implicit resolution in the time domain and no additional window functions are necessary.
- WT uses different dilations (scales) of the mother wavelet which provides better time localization of higher frequencies. For example, in Figure 5 it is apparent that the dilation at scale $a = 1$ which corresponds to the highest frequency, also has the narrowest support interval.
- Various wavelets exist and new ones can be designed so the most appropriate wavelet family can be selected depending on the problem at hand.
- Wavelets have an implicit relation to filters and filter banks, so the software implementation of Discrete Wavelet Transform (DWT) has relatively low complexity and high performance.

## 3.2.1. Continuous Wavelet Transform

Continuous Wavelet Transform (CWT) is most often used to analyze wavelet properties which cannot be captured and described easily if working directly with discrete transformations. Many of these analytical properties can be transferred and used for the discretized versions of the wavelets. We mention CWT here only to highlight its similarities to Fourier Transform and describe its relation to the Discrete Wavelet Transform.

Given a mother wavelet $\psi(t)$ and a function $f(t) \in L^2(\mathbb{R})$, the wavelet coefficients are defined as:

$$\tilde{f}(a,b) = \langle f, \psi_{a,b} \rangle = \int f(t)\psi_{a,b}(t)\, dt$$

where

$$\psi_{a,b}(t) = \frac{1}{\sqrt{a}}\psi\left(\frac{t-b}{a}\right), a \in \mathbb{R}^+, b \in \mathbb{R}$$

are normalized dilations and translations of the mother wavelet.

Similarly to the Fourier transform, CWT decomposes the function $f(t)$ in a basis of $L^2(\mathbb{R})$. The basis functions, however, are localized. The size and location of the support of each basis function depends on the dilation and translation parameters $a$ and $b$. Each wavelet coefficient $\tilde{f}(a,b)$ measures the correlation (similarity) between $f$ and $\psi_{a,b}$.

### 3.2.1.1. Convolution

CWT can also be expressed as a convolution. In the following sections we will see how this representation leads to the relationship between DWT and filter banks.

Let us define the function $\bar{\psi}_a(t)$:

$$\bar{\psi}_a(t) = \frac{1}{\sqrt{a}}\psi(\frac{-t}{a}) = \psi_{a,b}(b-t)$$

With this definition we obtain:

$$\tilde{f}(a,b) = \int f(t)\bar{\psi}_a(b-t)dt = (f * \bar{\psi}_a)(b),$$

where $*$ is the convolution operator.

### 3.2.1.2. Scaling Function and Signal Reconstruction

The reconstruction formula for the wavelet transform is the following:

$$f(t) = \frac{1}{C_\psi} \int_0^{+\infty} \int_{-\infty}^{+\infty} \tilde{f}(a,b) \psi_{a,b}(t) db \frac{da}{a^2},$$

where $C_\psi$ is a constant determined by the wavelet.

> ***Intuition:*** This formula essentially combines all basis functions $\psi_{a,b}(t)$ (the translations and dilations of the mother wavelet) and multiplies them by the wavelet coefficients to reconstruct the original function. This is analogous to how we typically represent a vector as a linear combination of basis vectors in a vector space.

In practice, we cannot calculate the coefficients for *all* scales $a$ up to infinity. If we knew the coefficients $\tilde{f}(a,b)$ only for $a < a_0$, then we would need some way to accumulate the extra information needed to reconstruct $f(t)$. This is achieved via an additional function, denoted $\phi(t)$ and called *scaling function*, also known as *father wavelet*. The scaling function is used in much the same way as the mother wavelet, but its dilations are interpreted differently. $\phi_{a_0,b} = a_0^{-1/2} \phi(a_0^{-1}(t-b))$ can be seen as a "cumulative term" which represents all dilations of the mother wavelet $\psi_{a,b}$ for $a \geq a_0$. Thus $f(t)$ can be reconstructed by using only $\phi_{a_0,b}$ and $\psi_{a,b}$ for $0 \leq a < a_0$. The exact details of the reconstruction formula using both mother and father wavelets, can be found in [4].

### 3.2.2. Wavelet Families

The first wavelet was constructed by Haar in 1910, nearly a century after the introduction of the Fourier transform. It was a piecewise constant function:

$$\psi(t) = \begin{cases} 1, & \text{if } 0 \leq t < 1/2 \\ -1, & \text{if } 1/2 \leq t < 1 \\ 0, & \text{otherwise} \end{cases}$$

The translations and dilations of this wavelet form an orthonormal basis. It is still being used as an "entry point" for experiments and education purposes due to its simplicity and orthonormality.

Several decades later, the work of Meyer resulted in the discovery of a family of orthonormal wavelet bases with functions $\psi$ which are infinitely continuously differentiable. This was the spark that triggered a widespread search for new orthonormal wavelet bases. The search culminated in the work of Belgian mathematician Ingrid Daubechies in 1988 [4]. Many of the current researches and experiments in the area of signal processing end up choosing one of the Daubechies wavelets as most appropriate for the task. She also developed two other families of wavelets called Symlets and Coiflets which have also proven to be effective. Symlets are modifications of the original Daubechies wavelets with increased symmetry. Coiflets are also near-symmetric and their scaling functions have vanishing moments (see Section 3.2.5).

### 3.2.3. Filters and Filter Banks

A *filter* is a device or process that removes some unwanted components or features from a signal [11]. This broad definition includes both physical devices (e.g. all kinds of electronic devices) as well as linear and non-linear mathematical processes that transform a signal in some way. For the practical purpose of processing an audio (or an image) signal, it makes sense to focus only on discrete linear time-invariant filters. Such filters have the following form:

$$\boldsymbol{y}[n] = \sum_{k} \boldsymbol{h}[k]\boldsymbol{x}[n-k] = \boldsymbol{h} * \boldsymbol{x},$$

where $\boldsymbol{x}[t] : \mathbb{Z} \to \mathbb{R}$ is a discrete-time signal, $\boldsymbol{h}[k] : \mathbb{Z} \to \mathbb{R}$ are the filter coefficients, $\boldsymbol{y}[t] : \mathbb{Z} \to \mathbb{R}$ is the filter response, and $*$ is the convolution operator. A special signal called an *impulse* plays an important role in filter analysis:

$$\boldsymbol{x}[t] = \begin{cases} 1, & \text{if } t = 0 \\ 0, & \text{otherwise} \end{cases}$$

The response of a filter to an impulse characterizes the filter itself. If the impulse response of a filter is finite, i.e., has a finite number of non-zero elements, the filter is called a *finite impulse response* (FIR) filter. This essentially means that $\boldsymbol{h}[t] = 0$ if $t$ is outside of a finite interval. Inversely, if the impulse response has infinitely many non-zero elements, then the filter is called an *infinite impulse response* (IIR) filter.

If $\boldsymbol{h}[t] = 0$ for all $t < 0$, then the filter is called *causal*. Causal filters do not require information "from the future", i.e., the calculation of $\boldsymbol{y}[n]$ involves only elements $\boldsymbol{x}[k], k \leq n$.

Causal FIR filters have non-zero coefficients only for $t \in [0, N]$ for some $N \in \mathbb{N}$. Thus, they can be represented by a finite vector $h \in \mathbb{R}^N$ and the application of the filter is the convolution of $h$ with the input signal $x$. In practice, $N$ is typically a reasonably small number, which makes such filters trivial to implement either as software procedures or as hardware circuits.

Filters can also be classified according to their frequency response. Some filters let only low frequencies pass through and block high frequencies. They are called *low-pass* filters. Similarly, *high-pass* filters pass through only high frequencies[3]. Such filters are often either not invertible or their inverse is an IIR filter. In both cases we do not have a practical procedure for reconstructing the original signal from the filter response. This may be a problem if our task involves not only analyzing the input signal, but also synthesizing a modified output signal.

It is possible to combine filters in so-called *filter banks* to achieve results which would be difficult or impossible using only a single filter. Applying *n* filters to the input signal produces an *n-channel filter bank*. In practice, 2-channel filter banks are most commonly used due to their simplicity and intrinsic relation to wavelets. Such filter banks are typically composed of one low-pass and one high-pass filter. If the filters are chosen properly, their responses can later be combined to reconstruct the original signal.

Filter banks which deconstruct the input by applying multiple filters to it are called *analysis banks*. Filter banks which take multiple inputs and combine them into one output are called *synthesis banks*. If we are given an analysis bank which decomposes a signal and would like to create a synthesis bank which reconstructs the original signal from the decomposition, the properties and structure of the synthesis bank will be tightly related to those of the analysis bank. Therefore analysis and synthesis banks are typically designed together and paired in larger filter banks as displayed in Figure 6.

Applying *n* filters to an input signal would produce *n* responses of the same length as the input and thus the output of the whole filter bank would be *n* times larger than the input. In order to preserve the length of the input, outputs of a filter bank are usually downsampled by keeping only every *n*-th component. In a 2-channel analysis bank the downsampling operator $\downarrow 2$ denotes the omission of every other element of the signal. During synthesis, the length needs to be restored. This is done by inserting zeros in the places of the missing components before applying the synthesis filters. This operation is denoted by the upsampling operator $\uparrow 2$. The downsampling and upsampling operations need to be taken into account when designing the filter bank. Indeed, applying these two operations consecutively to a signal would result in setting every other element of the signal to zero. If the filter bank is not designed to handle it, this could result in loss of information.

---

[3] Since the frequencies of discrete signals range from 0 to π, "low frequencies" in this context means "frequencies near zero", and "high frequencies" means "frequencies near π".

Figure 6 shows examples of how filter banks are typically visualized.

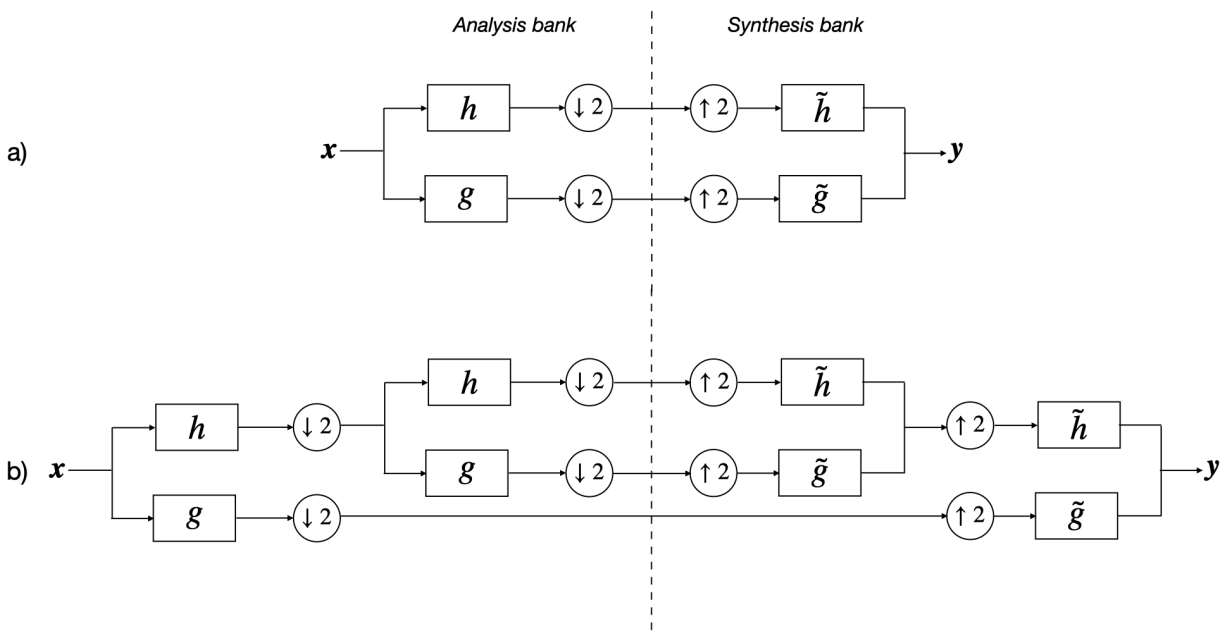

*Figure 6. a) Analysis and synthesis filter banks composed of two filters $h$ and $g$ and their inverses $\tilde{h}$ and $\tilde{g}$. The dashed line in the middle represents the point where the decomposition of the signal is typically transformed, transported, or analysed. b) The same filter bank as in a) is applied recursively to the output of its low-pass filter $h$. Note how the synthesis bank needs to mirror the structure of the analysis bank.*

One desirable property of filter banks is *perfect reconstruction (PR),* meaning that the filter bank accurately reconstructs the input signal after decomposing it. The analysis bank of a PR filter bank is invertible and the synthesis bank is its inverse. Theoretically there are two conditions which describe the perfect reconstruction property. The mathematical formulation of these conditions involves representing the filter bank in the so-called *z-domain* using the $\mathcal{Z}$-transform:

$$\mathcal{Z}\{\boldsymbol{h}[t]\}(z) = \sum_t \boldsymbol{h}[t] z^{-t},$$

where, in general, $z$ is a complex number. In the z-domain the application of a filter $\boldsymbol{h}$ is represented as polynomial multiplication:

$$Y(z) = H(z)X(z),$$

where $H$, $X$ and $Y$ are the $\mathcal{Z}$-transforms of the filter $\boldsymbol{h}$, the input signal $\boldsymbol{x}$ and the filter response $\boldsymbol{y}$ respectively. Thus we can represent the transformation applied by the filter bank in Figure 6.a) as:

$$Y(z) = \tilde{H}(z)(\uparrow 2)(\downarrow 2)H(z)X(z) + \tilde{G}(z)(\uparrow 2)(\downarrow 2)G(z)X(z),$$

where $\tilde{H}$, $H$, $\tilde{G}$, $G$, $X$ and $Y$ are the $\mathcal{Z}$-transforms of the filters $\tilde{\boldsymbol{h}}$, $\boldsymbol{h}$, $\tilde{\boldsymbol{g}}$, $\boldsymbol{g}$, and the input and output signals $\boldsymbol{x}$ and $\boldsymbol{y}$ respectively. As mentioned above, the consecutive application of the downsampling and upsampling operators sets every other element of the signal to zero. In the z-domain this can be written as:

$$(\uparrow 2)(\downarrow 2)H(z)X(z) = \frac{1}{2}(H(z)X(z) + H(-z)X(-z))$$

This is an even function where every odd power of $z$ is canceled. After applying this formula and rearranging the terms in the previous equation, we obtain:

$$Y(z) = \frac{1}{2}(\tilde{H}(z)H(z) + \tilde{G}(z)G(z))X(z) + \frac{1}{2}(\tilde{H}(z)H(-z) + \tilde{G}(z)G(-z))X(-z)$$

In a PR filter bank, the output is equal to the input, i.e., $Y(z) = X(z)$. If we would like to allow the filter bank to introduce a time delay of $l$ time steps, we could generalize this requirement as $Y(z) = z^{-l}X(z)$. Thus, from the formula above we obtain the two conditions under which the filter bank would have perfect reconstruction:

$$\tilde{H}(z)H(z) + \tilde{G}(z)G(z) = 2z^{-l}$$

and

$$\tilde{H}(z)H(-z) + \tilde{G}(z)G(-z) = 0$$

The former condition is known as the *no distortion* requirement and the latter is known as the *alias cancellation* requirement.

Working "backwards" from these two conditions it is possible to devise procedures for creating new 2-channel PR FIR filter banks. Historically there have been several approaches to building such filter banks. Designers usually start from the low-pass analysis filter. The high-pass filter is then constructed by alternating the signs and optionally flipping the order of the low-pass filter coefficients: $g_i = (-1)^i h_i$ or $g_i = (-1)^i h_{N-i}$. The filter pairs constructed in such a manner are known as quadrature mirror filters (QMF) and conjugate quadrature filters (CQF) respectively. These constructions are neither required nor sufficient to produce a PR filter bank, but they illustrate the simplicity of steps which are typically taken to construct such filters. Further details about the formal conditions which a PR filter bank must meet, as well as some efficient methods for constructing new filter banks, can be found in [12].

## 3.2.4. Discrete Wavelet Transform

Alfréd Haar was the first to apply a discrete wavelet transform (DWT) in the beginning of the 20th century, although he did not use this terminology at the time. Several decades later the relationship between DWT and filter banks was noticed as the area was starting to gain popularity and attract researchers. In many present-day texts on the topic, filter banks are actually used to *define* DWT.

Probably the most intuitive way to understand DWT is by analogy with CWT. If we consider a discretely sampled version of a wavelet, then the convolutional form of CWT would be analogous to the application of a linear time-invariant filter:

$$\bm{y}[n] = \bm{w} * \bm{x} = \sum_k \bm{w}[k]\bm{x}[n-k],$$

where $\bm{w} : \mathbb{Z} \to \mathbb{R}$ are the wavelet coefficients. The scaling function of the wavelet can also be represented as a filter. Due to their mathematical properties, it can be shown that the scaling function corresponds to a low-pass filter and the wavelet corresponds to a high-pass filter. Thus we can create a 2-channel filter bank. If the wavelet filter is applied recursively to the output of the scaling filter, it would produce the coefficients for the next dilation of the mother wavelet. Thus the filter bank can be recursively cascaded multiple times to get as many levels of wavelet coefficients as necessary. This process is shown in Figure 7. The responses of the wavelet and scaling filters at each level are often called *detail* and *approximation* respectively.

> ***Intuition:*** Notice the difference in the role of the scaling function (or scaling filter respectively) in CWT and DWT. In CWT all wavelet coefficients up to some arbitrary scale $a_0$ can be calculated directly from the formula $\tilde{f}(a,b) = \langle f, \psi_{a,b} \rangle$ without the use of the scaling function $\phi$. The scaling function is then used to accumulate the information for wavelet scales larger than $a_0$. In DWT the semantic of the scaling filter is analogous but it actually participates in the calculation of wavelet coefficients in all levels after the first. So the calculation of wavelet coefficients for level $l > 1$ is performed by $l - 1$ consecutive applications of the scaling (low-pass) filter, followed by one application of the wavelet (high-pass) filter.

This filter bank construction of DWT leads to a special scale where the scaling parameter $a = 2^{l-1}$ at level $l$. It is known as *dyadic* or *octave* scale. In this scale every dilation of the wavelet (and the scaling function) has a center frequency which is half the frequency of the previous level, i.e., one octave lower. It also produces half the number of output coefficients which is obvious from the downsampling operators in Figure 7.

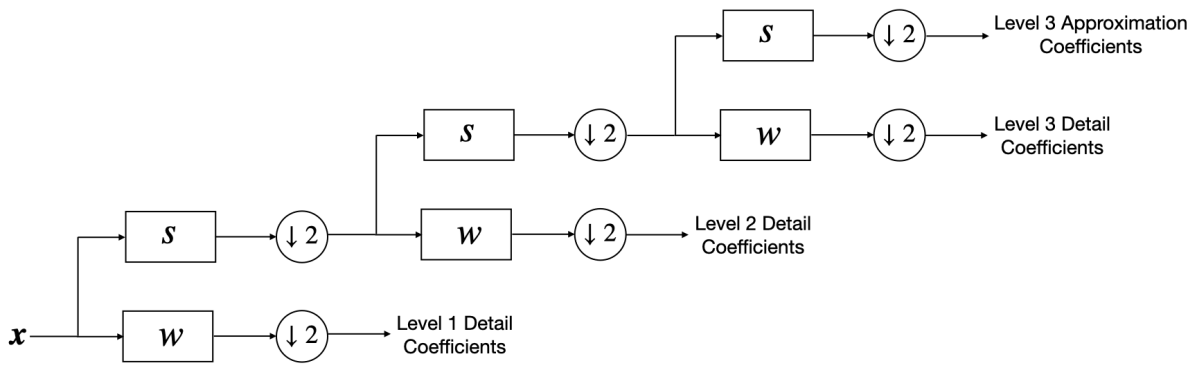

*Figure 7. Discrete Wavelet Transform represented as a filter bank. The low-pass filter $s$ corresponds to a scaling function and the high-pass filter $w$ corresponds to a wavelet.*

It can be seen in Figure 7 that the number of detail coefficients is halved at each following level of decomposition. Since the input to each successive level is the low-pass output of the previous level, this means that higher-level detail coefficients correspond to lower frequencies. In other words, we obtain fewer coefficients for lower frequencies and more coefficients for higher frequencies. This is an essential difference between DWT and STFT and it is often represented via diagrams similar to the one in Figure 8.

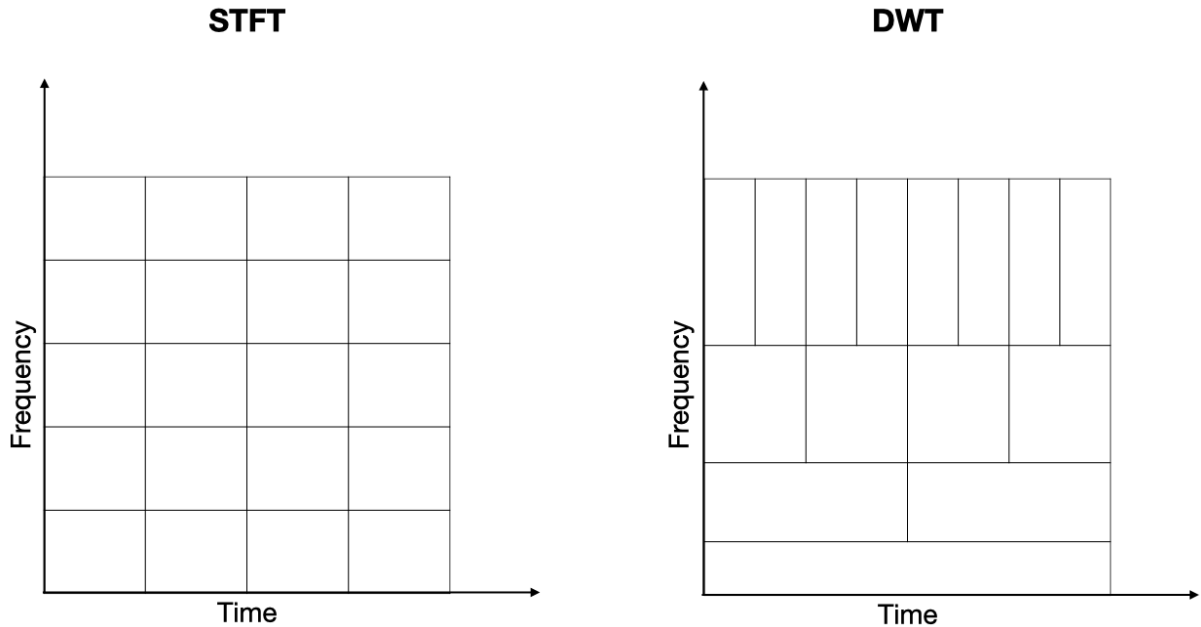

*Figure 8. Frequency and time resolution of STFT compared to DWT*

***Highlight:*** The relationship between the frequency and the wavelength of a wave is reciprocal. The coefficient structure of DWT follows this same pattern and so it seems to be a more natural representation of signals in this sense. This irregular structure, however, poses some implementation difficulties.

Most machine learning frameworks, such as TensorFlow, are designed to work only with rectangular matrices, especially for computations powered by GPUs. Therefore, in order to use the DWT coefficients in an ML model, we need to restructure them in some way. The easiest option is to arrange them in a one-dimensional vector as displayed in Figure 9. This method has the disadvantage of discarding the time adjacency of coefficients of different levels. Notice, for example, the coefficients marked with white dots in the figure. They represent phenomena which occurred at the same time, but in the flattened representation this information is discarded. There may be other ways to restructure the DWT coefficients, such as duplicating them or filling the rows with zeros so that the matrix becomes rectangular, but each of these methods has drawbacks which need to be considered with regards to how they will affect the chosen ML model.

> One modification of DWT can be used to mitigate the problem with the irregular coefficient structure. It is called Wavelet Packets and is described in detail in a separate section below.
>
> 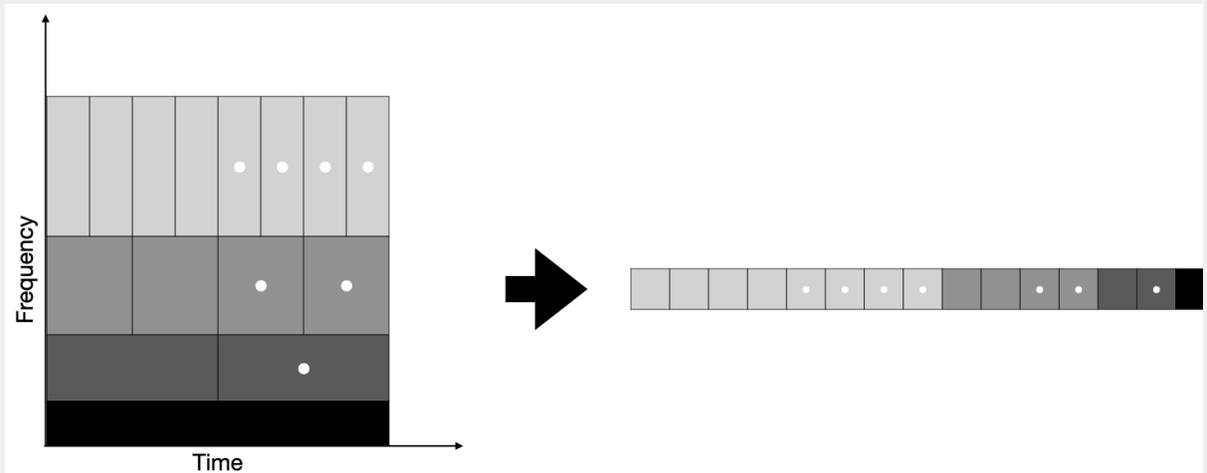
>
> *Figure 9. Flattening of DWT coefficients. Some coefficients are marked with white dots to highlight their time adjacency in both arrangements.*

The wavelets used in DWT often do not have analytic forms and their sampled versions are not readily available. The construction of such wavelets starts from the scaling filter of the filter bank. The filter bank is then built to have perfect reconstruction and it is often designed to be orthogonal or biorthogonal. The equations which relate the filters $\boldsymbol{s}$ and $\boldsymbol{w}$ to the continuous-time scaling and wavelet functions $\phi$ and $\psi$ are:

$$\phi(t) = \sqrt{2} \sum_k \boldsymbol{s}(k)\phi(2t - k)$$

$$\psi(t) = \sqrt{2} \sum_k \boldsymbol{w}(k)\phi(2t - k)$$

These equations can be used to transfer properties of the filters to properties of the continuous-time functions and vice-versa. Additionally, procedures have been developed, with which one can generate candidate filters and verify whether they meet the design goals using the above equations. Wavelet design goals depend on the purpose for which the wavelet is being constructed but they often include properties such as orthogonality, orthonormality, biorthogonality, symmetry of coefficients, or a particular number of vanishing moments (described in the following section).

> ***Highlight:*** Discrete wavelets are represented and constructed as 2-channel filter banks. The analytical forms of their wavelet and scaling functions are typically not available.

## 3.2.5. Vanishing Moments

By definition, the *k-th moment of a function* $f(x)$ is the integral

$$\int_{-\infty}^{\infty} x^k f(x) dx$$

A wavelet $\psi(t)$ is said to have $p$ *vanishing moments* if its first $p$ moments are zero:

$$\int t^k \psi(t) dt = 0 \text{ for } 0 \leq k < p$$

A wavelet with $p$ vanishing moments is orthogonal to polynomials of degree less than $p$. This means that if we perform CWT on such a polynomial, the first-level wavelet coefficients will be zero and the polynomial will be approximated entirely by the first-level scaling function. Analogously, in the discrete case, only the low-pass filter would produce nonzero coefficients.

Intuitively, the more vanishing moments a wavelet has, the better it can approximate more complicated signals. However, there is a tradeoff to this property. A theorem due to Ingrid Daubechies states that a wavelet $\psi$ with $p$ vanishing moments that generates an orthonormal basis of $L^2(\mathbb{R})$ has a support size larger than or equal to $2p-1$. Another theorem adds that filters corresponding to such wavelets have at least $2p$ nonzero coefficients. In other words, the number of vanishing moments establishes a lower bound for the size of the support of the wavelet and the size of the corresponding filter bank respectively. In fact, Ingrid Daubechies also proved that the Daubechies family of wavelets have minimal filter banks in this sense, i.e., a Daubechies wavelet with $p$ vanishing moments has filters with exactly $2p$ coefficients [4].

The large number of coefficients in a wavelet filter bank can clearly affect performance (there are more things to compute) but it also has another disadvantage. Singularities in the input signal, such as spontaneous bursts or interruptions, can cause the corresponding wavelet coefficients to become large. For example, if we have a singularity at time $t_0$ and a wavelet filter with $K$ coefficients, then at each scale there will be $K$ translations of the mother wavelet which include $t_0$. Figure 1 can be used as a visualization of this phenomenon. Indeed, we can assume $h$ in the figure to be a wavelet (high-pass) filter of length $K = n$, and $x_t$ to be a singularity, say a spontaneous burst where $x_t$ is much larger than $x_{t-1}$ and $x_{t+1}$. $x_t$ participates in the calculation of $y_t, y_{t+1}, \ldots, y_{t+K-1}$, so a total of $K$ wavelet coefficients will be affected by this singularity. Therefore, if we are working with signals with high density of singularities, it is better to choose a wavelet with fewer vanishing moments and smaller filter size respectively.

> *Highlight:* Wavelets with more vanishing moments can approximate complicated signals more efficiently. However, their filters have more coefficients and thus are more difficult to compute and more sensitive to bursts and irregularities in the signal.

### 3.2.6. Multiresolution Analysis

Multiresolution analysis (MRA) is a concept borrowed from the field of image processing and adapted to wavelet theory by Mallat and Meyer [4]. Analyzing wavelets through MRA has led to many significant results such as the definition of *conjugate mirror filters* – a powerful type of filters used to construct orthogonal wavelet filter banks, including the celebrated Daubechies wavelets. MRA also provides a foundation for defining the concept of wavelet packets which is the topic of the following section.

The idea of MRA is that an object of interest, such as a signal $f \in L^2(\mathbb{R})$, can be analyzed at different levels, or *resolutions*. The resolutions are represented by closed nested subspaces $\cdots \supset V_{-1} \supset V_0 \supset V_1 \supset \cdots$ of $L^2(\mathbb{R})$. A system of such subspaces needs to satisfy certain conditions in order to be considered a *multiresolution approximation*. These conditions are described in detail in [4]. Then for each space $V_j$ we can find an approximation of $f$ as an *orthogonal projection* on $V_j$ which is the function $f_j \in V_j$ that minimizes $\|f - f_j\|$.

A theorem states that the dilations and translations $\phi_{a,b}$ of a scaling function $\phi$ matching certain criteria, form orthonormal bases of a multiresolution approximation.

In particular, the set of translations $\{\phi_{a,b}\}_{b\in\mathbb{Z}}$ at a given scale $a$ is an orthonormal basis of $V_a$. Thus, scaling functions can be used for multiresolution analysis. This is the reason why it is often considered that an orthogonal wavelet is uniquely defined by its scaling function $\phi$, or scaling filter $s$ in the discrete case, and the construction of such wavelets often starts from $\phi$ or $s$ respectively.

The role of wavelet functions in MRA is that they provide the details at each resolution. Let us consider two adjacent resolutions $V_a$ and $V_{a-1}$. By definition we have $V_a \subset V_{a-1}$. Let $W_a$ be the orthogonal complement of $V_a$ in $V_{a-1}$:

$$V_{a-1} = V_a \oplus W_a$$

Mallat and Meyer proved that orthonormal bases of the spaces $\{W_a\}_{a\in\mathbb{Z}}$ can be constructed from the translations and dilations $\psi_{a,b}$ of a wavelet $\psi$ related to the scaling function $\phi$ that generates the multiresolution approximation $\{V_a\}_{a\in\mathbb{Z}}$. Thus, if the approximations of a function $f$ at resolutions $V_a$ and $V_{a-1}$ are $f_a$ and $f_{a-1}$ respectively, and if $\tilde{f}_a$ is the projection of $f$ on $W_a$, then:

$$f_{a-1} = f_a + \tilde{f}_a$$

So the wavelet coefficients $\tilde{f}_a$ provide the details of $f$ that are present at resolution $V_{a-1}$ but disappear at resolution $V_a$. (Hence, the spaces $\{W_a\}_{a\in\mathbb{Z}}$ are also known as *detail spaces*.) Graphically this structure can be represented as in Figure 10. It is no coincidence that Figure 10 resembles the tree structure of DWT in Figure 7. These are essentially two different representations of the same decomposition. In fact, Lemarié proved that *all* orthonormal wavelet bases have a corresponding multiresolution approximation.

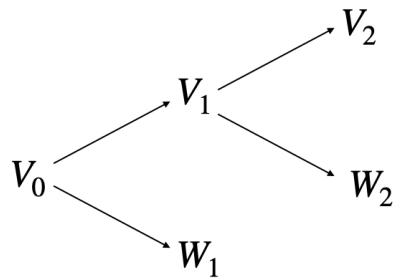

*Figure 10. Structure of a multiresolution approximation*

### 3.2.7. Wavelet Packets

In a multiresolution approximation $\{V_a\}_{a\in\mathbb{Z}}$ with detail spaces $\{W_a\}_{a\in\mathbb{Z}}$ we saw that the space $V_{a-1}$ can be decomposed into subspaces $V_a$ and $W_a$: $V_{a-1} = V_a \oplus W_a$. Put differently, for each $a \in \mathbb{Z}$, the basis $\{\phi_{a-1,b}\}_{b\in\mathbb{Z}}$ of $V_{a-1}$ can be decomposed into two orthonormal bases: $\{\phi_{a,b}\}_{b\in\mathbb{Z}}$ of $V_a$ and $\{\psi_{a,b}\}_{b\in\mathbb{Z}}$ of $W_a$. A theorem due to Coifman, Meyer, and Wickerhauser states that such decomposition can be performed in *any* space $U_a \subset L^2(\mathbb{R})$ that has an orthonormal basis constructed from the translations of a function on a dyadic scale. In particular, this basis decomposition can also be applied in the detail spaces $\{W_a\}_{a\in\mathbb{Z}}$.

Since the spaces $\{V_a\}_{a\in\mathbb{Z}}$ and $\{W_a\}_{a\in\mathbb{Z}}$ will be treated identically, let us denote for convenience $U_a^0 := V_a$ and $U_a^1 := W_a$. Then the space $U_{a-1}^j$ can be decomposed by:

$$U_{a-1}^j = U_a^{2j} \oplus U_a^{2j+1}$$

The generalized spaces $U_a^j$ are known as *wavelet packet spaces*. Wavelet packets allow us to build wider MRA decomposition trees as shown in Figure 11.

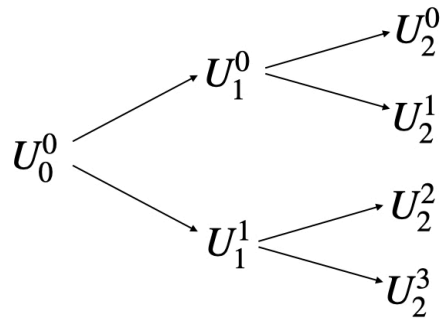

*Figure 11. Binary tree of a multiresolution approximation with wavelet packet spaces*

We are not required to decompose all of the spaces $U_a^j$ at each level. If we choose and decompose only some of the spaces, we would still be able to reconstruct the input. Thus we can compose arbitrarily shaped and potentially unbalanced wavelet packet trees.

The decomposition of wavelet packet bases can be performed by applying a pair of conjugate mirror filters to the basis functions – the same type of filters that are commonly used in DWT. This allows us to transfer the concept of wavelet packets directly onto DWT filter banks and construct analogous wider banks as shown in Figure 12. The resulting transformation is known as *Wavelet Packet Transform* (WPT).

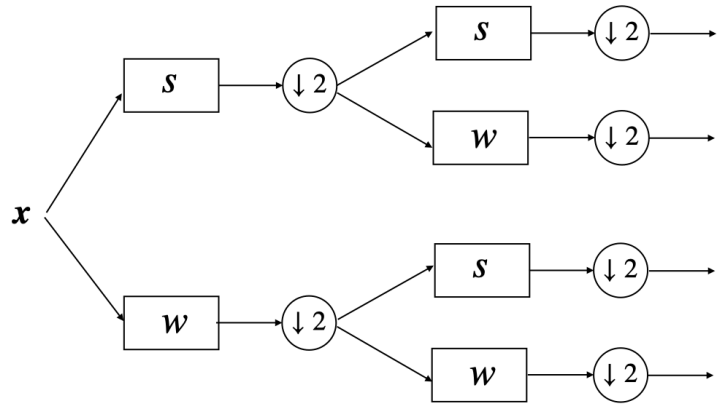

*Figure 12. Wavelet Packet Transform*

*Highlight:* Practically, if we ignore the MRA-based definition of wavelet packets, we can view WPT as a simple extension of DWT where we apply the filter bank recursively not only to the approximation but also to the detail coefficients. The total number of coefficients remains the same but they are rearranged into different frequency bands. This fact has two implied benefits.

Firstly, we obtain a sub-octave frequency resolution. This is useful when dealing with speech signals since human voice frequencies are mostly concentrated in one or two octaves. For example, the voices of two adult males often have similar base frequencies.

Secondly, by using a full binary tree of wavelet packet spaces we can construct rectangular coefficient matrices where we have the same number of coefficients for each frequency band. This simplifies the usage of WPT output as input for various machine learning models, as mentioned in a previous section.

The disadvantage of WPT is that it can become many times slower to compute than DWT depending on the shape of the wavelet packet tree. This is simply because in WPT we need to apply the same filters as in DWT, but more times.

### 3.2.7.1. Frequency Ordering

As mentioned in previous sections, in DWT the scaling filter $s$ is a low-pass filter and the wavelet filter $w$ is a high-pass filter, so they split the signal into low and high frequencies. Additionally, the dyadic scale which is used for translating and dilating the functions at different levels results in lowering the center frequency of the output at each successive level by one octave. This makes it easy to reason about the frequency band to which the coefficients at each scale correspond.

However, frequency ordering in WPT is not as trivial to describe. Instead of only "low-pass" and "high-pass" output at level $a$, we have $2^a$ different outputs $j = 0, 1, ..., 2^a - 1$ corresponding to different frequency bands. Coifman and Wickerhauser proved that a permutation $k = G[j]$ of the index $j$ that arranges the outputs in increasing order of center frequencies, satisfies the following properties:

$$G[2j] = \begin{cases} 2G[j], & \text{if } G[j] \text{ is even} \\ 2G[j] + 1, & \text{if } G[j] \text{ is odd} \end{cases}$$

$$G[2j+1] = \begin{cases} 2G[j] + 1, & \text{if } G[j] \text{ is even} \\ 2G[j], & \text{if } G[j] \text{ is odd} \end{cases}$$

The inverse of this permutation $j = G^{-1}[k]$ is known in coding theory as *Gray code* after physicist Frank Gray.

> **Highlight:** Gray code can be used to obtain the correct frequency order of WPT coefficients.

### 3.2.8. Applications

One of the most commonly used Python packages for applying wavelet transformations is PyWavelets, often abbreviated as `pywt`. It contains functions for working with CWT, DWT, and WPT.

### 3.2.8.1. CWT vs DWT

In practice, CWT is performed by numerical approximation of the integrals used to calculate the wavelet coefficients. The dilation and translation parameters are discretized, so the transformation is not entirely continuous (if it were, it would produce an infinite number of coefficients), but it still uses the analytical forms of the wavelet and scaling functions.

DWT on the other hand, is calculated using filter banks. This method is much more efficient, but it also has some limitations. For example, the dyadic scale is often built into the wavelet filter bank itself, so we cannot specify arbitrary dilation factors without modifying the filter bank. Also, many discrete wavelets do not have analytical representations at all, so it may be difficult to reason about the wavelet properties in some cases, or even to plot what the wavelet function looks like.

As of this writing, the PyWavelets library does not support any wavelet families which can be used for both CWT and DWT, i.e., all wavelets are either discrete (defined as filter banks) or continuous (defined as analytical functions). This fact has little if any practical implications, however, since it is rare for a practical task to require the application of both CWT and DWT with exactly the same wavelet family.

### 3.2.8.2. Frequency Mapping

Wavelets are functions with finite support, so there is no direct mapping between a wavelet and a specific linear frequency in the way a sinusoid, for example, can be mapped to its frequency. However, a mother wavelet can still be approximated with a sinusoid in its support interval and its *central frequency* can be evaluated in this way. This is the frequency which the mother wavelet most closely resembles. If we know the central frequency of a wavelet, calculating the frequencies to which its dilations correspond is easy, because the frequency is inversely proportional to the dilation factor.

For example, for a discrete filter with a dyadic scale, the scaling factor at level $l$ is $a = 2^{l-1}$. This means that the wavelet at scale $a + 1$ is twice as wide as it is at scale $a$. Twice larger wavelength corresponds to twice smaller frequency, so the corresponding frequency is halved at each level. Thus, the wavelet at level $1$ has the highest frequency which becomes $2^5 = 32$ times smaller at dilation level $6$, for example.

The central frequency of a continuous wavelet can be obtained analytically, but for discrete wavelets it can be non-trivial to calculate. PyWavelets contains a function

**central_frequency** which can be used to obtain it. The package also offers a function **scale2frequency** which can be used to calculate the matching frequency at a given scale of a given wavelet.

### 3.2.8.3. Scaleograms

Scaleograms, similarly to spectrograms, are heat maps of wavelet coefficients. The main difference between scaleograms and spectrograms is that along the *y* axis scaleograms contain the dilation levels (scales) of the wavelet instead of the frequency spectrum. A sample scaleogram is shown in Figure 13 using the same audio recording as the one used in the spectrogram example in Figure 4. The code used to plot this scaleogram is given in Listing 3. Note that Listing 3 also contains an example of how to apply DWT in practice.

*Listing 3. Plotting a scaleogram. Complete implementation can be found at:*
*https://github.com/rganchev/speech-signal-processing-for-ml*

```python
import matplotlib.pyplot as plt
import numpy as np
import pywt

def plot_scaleogram(ax, wav):
  levels = 6 # use six levels of decomposition
  wavelet = 'sym8' # use a Symlet wavelet with 8 vanishing moments
  # Use wavedec to apply several levels of decomposition at once.
  coeffs = pywt.wavedec(wav, wavelet, level=levels, mode='periodization')
  # wavedec returns the coeffs in reverse order (higher scales first).
  # Flip them, so we get a better ordering in the heat map.
  coeffs = np.flip(coeffs, axis=0)
  # The first level coefficient vector has the largest size.
  size = coeffs[0].shape[0]
  # Repeat the coefficients so that we obtain a rectangular matrix.
  arr = np.array([np.repeat(c, size / c.shape[0]) for c in coeffs])
  # Plot the heatmap using plt.imshow
  ax.imshow(np.abs(arr), aspect='auto')
  ax.set(title='Scaleogram', xlabel='translation', ylabel='scale',
         yticks=range(0, levels + 1, 2),
         yticklabels=range(1, levels + 2, 2))

fs, wav = read_wav_file() # from Listing 1
wav = wav[:65536]
```

```
_, (ax1, ax2) = plt.subplots(nrows=2)
plot_waveform(ax1, wav) # from Listing 2
plot_scaleogram(ax2, wav)
```

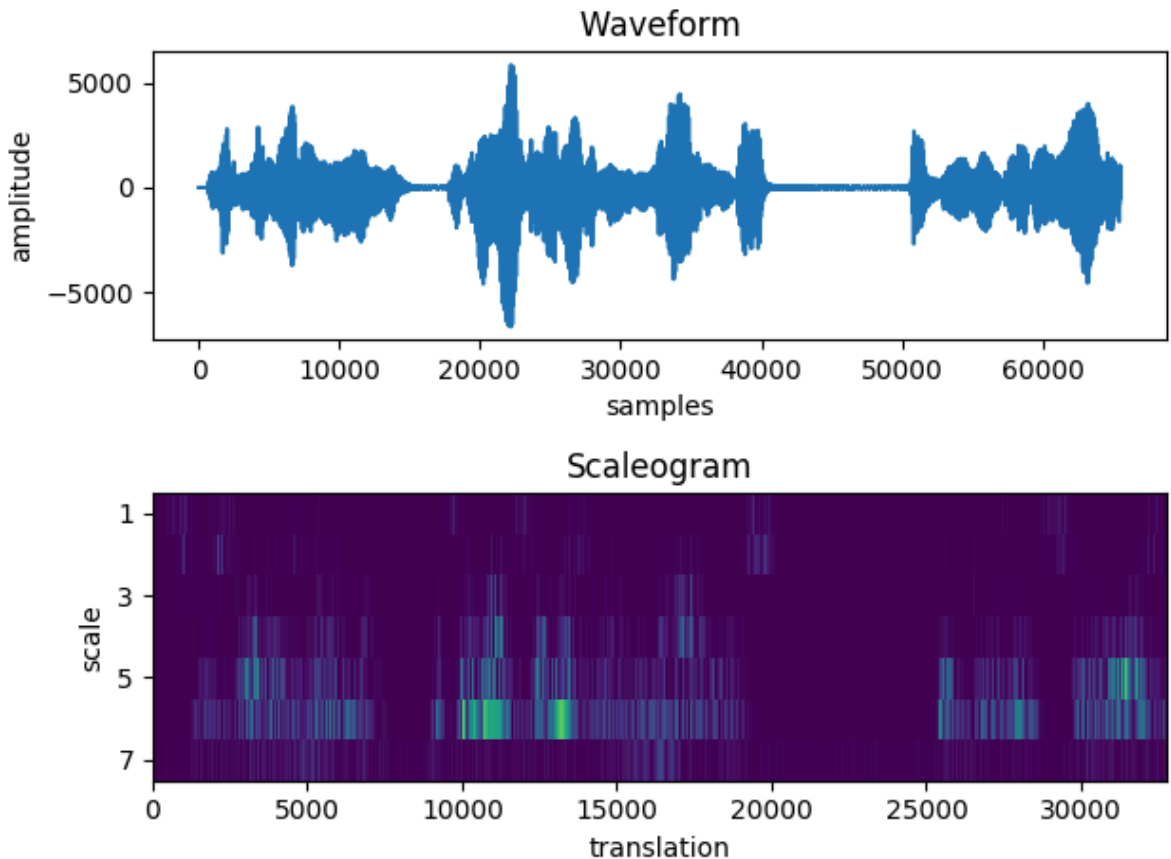

*Figure 13. A waveform sampled at $16kHz$ and its corresponding scaleogram. The wavelet used for the decomposition is a Symlet with 8 vanishing moments. Six levels of decomposition were applied. The bottommost row of the scaleogram represents the approximation coefficients at level 6 and the six rows above represent the detail coefficients at levels 1–6. Coefficients at levels 2–6 were repeated multiple times in order to arrange a rectangular matrix.*

The example in Figure 13 uses a Symlet wavelet with 8 vanishing moments. Its approximate central frequency is $0.666$ oscillations per sample. At a $16kHz$ sampling rate we obtain a $0.666 * 16 = 10.656 kHz$ central linear frequency of the mother wavelet (i.e., the wavelet at scale $a = 1$). This means that at levels 5 and 6 the corresponding frequencies are $10656/2^4 = 666 Hz$ and $10656/2^5 = 333 Hz$ respectively. These are the scales at which we see the brightest areas in the scaleogram. These frequency bands also approximately correspond to the brightest areas in the spectrogram in Figure 4.

Scaleograms can also be plotted for Wavelet Packet Transforms. If we perform complete decomposition of the wavelet packet tree, we will obtain a rectangular matrix of coefficients so no need for further adjustments would be necessary. An example of using WPT and plotting its scaleogram is shown in Listing 4 and Figure 14 respectively.

*Listing 4. Plotting a WPT scaleogram. Complete implementation can be found at: https://github.com/rganchev/speech-signal-processing-for-ml*

```python
import matplotlib.pyplot as plt
import numpy as np
import pywt

def plot_wpt_scaleogram(ax, wav):
  L = 6
  wp = pywt.WaveletPacket(data=wav, wavelet='sym8', maxlevel=L)
  # Stack the coefficients of all nodes in a rectangular matrix.
  # Use order='freq' to order the coefficients by frequency band.
  coeffs = np.stack([node.data for node in wp.get_level(L, order='freq')])
  ax.imshow(np.abs(coeffs), aspect='auto', origin='lower')
  ax.set(title='Scaleogram', xlabel='translation', ylabel='scale')

fs, wav = read_wav_file() # from listing 1
wav = wav[:65536]

_, (ax1, ax2) = plt.subplots(nrows=2)
plot_waveform(ax1, wav) # from listing 2
plot_wpt_scaleogram(ax2, wav)
```

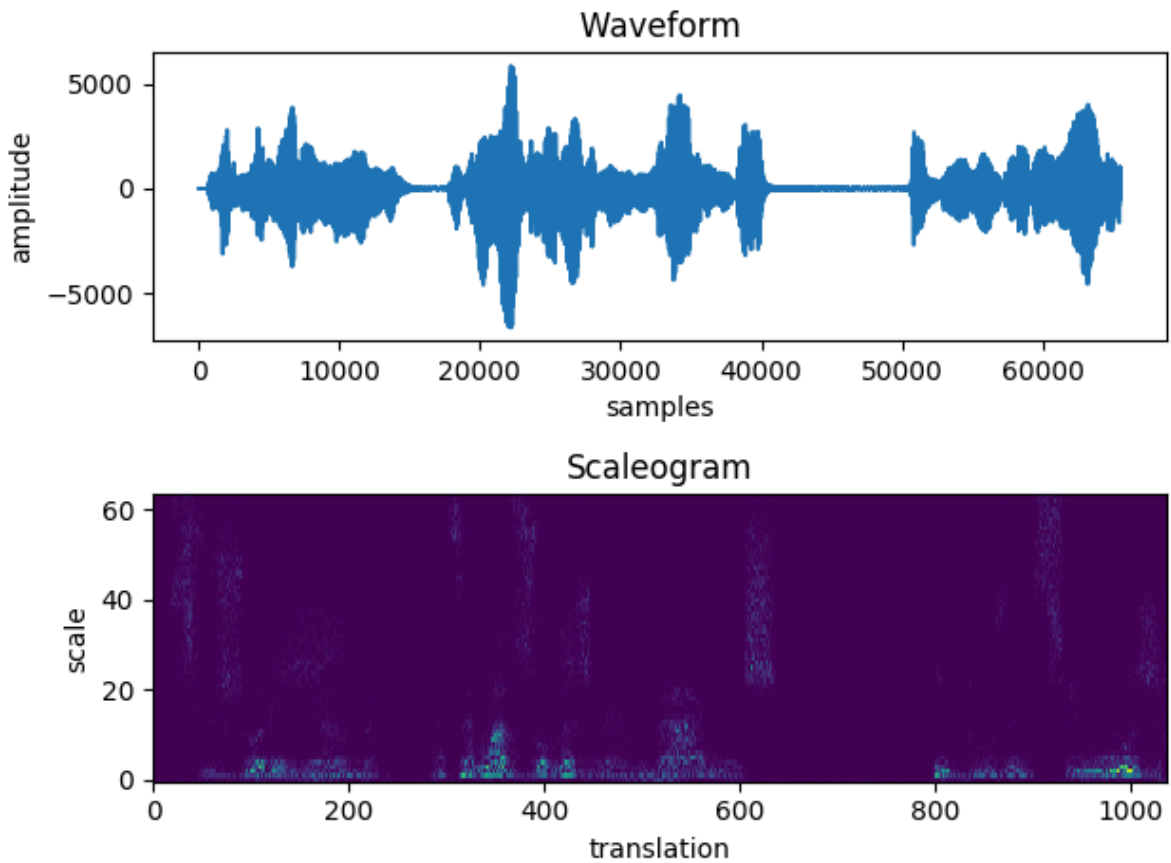

*Figure 14. A waveform and its corresponding WPT scaleogram. The wavelet used for the decomposition is a Symlet with 8 vanishing moments. Six levels of decomposition were applied which have produced $2^6 = 64$ leaf nodes in the wavelet packet tree.*

The following facts can be observed in the WPT scaleogram in Figure 14:
- Even though the same number of decomposition levels were used as in the DWT scaleogram in Figure 13, WPT has generated a lot more coefficients for each time step. Thus WPT has a finer frequency resolution than DWT.
- Figure 14 resembles the spectrogram in Figure 4. Unlike the spectrogram, however, most of the energy (bright pixels) in the scaleogram is concentrated in fewer areas and not scattered across the whole spectrum.

# 4. Speech Intelligibility Metrics

A multitude of metrics have been developed for evaluating the quality of an approximated speech signal $\hat{s}$ compared to a reference signal $s$. Such metrics are often referred to as "relative" as opposed to the "absolute" metrics which evaluate the quality of $\hat{s}$ on its own, without the presence of a reference.

Many of the available relative metrics, however, are difficult or impossible to differentiate. Therefore, simple metrics such as mean squared error (MSE) or signal-to-noise ratio (SNR) remain some of the most widely used functions for calculating the loss in statistical inference models for speech separation. MSE or SNR may lead to decent results when used as loss functions but they are flawed when comparing the performance of two different approximations. One of the problems they have is that they penalize transformations of the signal that do not affect the intelligibility of speech. For example, consider an approximation of $s$ which applies a time-invariant gain $\alpha \neq 1$: $\hat{s} = \alpha s$. Then:

$$MSE(s, \hat{s}) := \frac{\|s - \hat{s}\|^2}{n} = \frac{\|s - \alpha s\|^2}{n} = \frac{(1-\alpha)^2}{n}\|s\|^2$$

and

$$SNR(s, \hat{s}) := 10 \log_{10}\left(\frac{\|s\|^2}{\|s - \hat{s}\|^2}\right) =$$
$$= 10 \log_{10}\left(\frac{\|s\|^2}{\|s - \alpha s\|^2}\right) = 10 \log_{10}\left(\frac{1}{(1-\alpha)^2}\right) = -20 \log_{10}|1-\alpha|$$

It can be seen from these equations that MSE and SNR vary significantly with the value of $\alpha$, but the gain only amplifies or diminishes the energy of the signal without modifying its frequency composition, so it does not actually affect intelligibility.

Time-invariant gains are not the only type of transformations that do not impact speech intelligibility significantly. In telephone systems, for example, low-energy static noises are a common occurrence. Band-pass filters are also often applied to cut-off frequency bands outside of the narrow-band voice frequency range of 300-3400 Hz in order to reduce the amount of data necessary to transmit the signal over a network [13]. Such transformations would also be penalized by MSE and SNR.

To mitigate such problems, in [14] it is suggested to break down the approximated signal $\hat{s}$ into different components[4]. Each of them is calculated using orthogonal projections of $\hat{s}$ onto vector spaces spawned by elements of the input signal. Thus we can isolate different error components such as interference, noise, and artifacts. More importantly, we can calculate the projections in such a way that they are not affected by a given set of "allowed distortions". Hence, the resulting metrics will not penalize such distortions.

The authors of [15] take this idea further and suggest a much simpler version of the metrics developed in [14]. It is referred to as "Scale-Invariant Signal-to-Distortion Ratio" (SI-SDR) and is described in detail below.

Some of the metrics mentioned so far put emphasis on intelligibility but are not specifically designed for speech. They can also work well with source separation tasks related to other types of audio signals such as music, for example. While being more general-purpose, such metrics do not make use of some of the specificities of speech signals.

The early progress of telecommunication systems has demanded the existence of accurate objective measures of speech quality. This has encouraged the development of speech-specific metrics such as PESQ (Perceptual Evaluation of Speech Quality) and STOI (Short-Time Objective Intelligibility), as well as their standardization by the International Telecommunications Union (ITU). Such metrics have later found their applications into modern machine-learning models for speech manipulation.

Even though newer and more accurate metrics have recently been developed and standardized by ITU, many of them are proprietary, so the freely available SI-SDR, PESQ and STOI are still some of the most widely used metrics in research papers on speech separation models. Each of these metrics is explained in detail below.

---

[4] The main metric developed in [14] is called SDR, abbreviated from Signal-to-Distortion Ratio or *Source*-to-Distortion Ratio. In this terminology "distortion" refers to the total error of the approximated signal represented as the sum of the three error components – interference, noise, and artifacts. Respectively, three other "sub"-metrics are also proposed – Source-to-Interference Ratio (SIR), Source-to-Noise Ratio (SNR, not to be confused with the other SNR metric described above), and Source-to-Artifacts Ratio (SAR). These metrics have a somewhat complicated mathematical formulation which will not be included here since it is not relevant to the current work.

## 4.1. Scale-Invariant Signal-to-Distortion Ratio (SI-SDR)

This metric, proposed in [15], is an improvement of SNR based on some ideas from the original SDR metric developed in [14]. In contrast to SDR, however, instead of making projections of $\hat{s}$ onto vector spaces spawned by input signal components, SI-SDR rescales either $s$ or $\hat{s}$ so that the residual vector $s - \hat{s}$ is orthogonal to $s$:

$$SI\text{-}SDR(s, \hat{s}) := 10 \log_{10} \left( \frac{\|\alpha s\|^2}{\|\alpha s - \hat{s}\|^2} \right)$$

where $\alpha s \perp \alpha s - \hat{s}$. The value of $\alpha$ can be calculated from the equation $\alpha s \cdot (\alpha s - \hat{s}) = 0$ where the operator $a \cdot b$ is the dot product of vectors $a$ and $b$. Indeed we have:

$$\alpha s \cdot (\alpha s - \hat{s}) = \alpha^2 \|s\|^2 - \alpha(s \cdot \hat{s}) = 0$$

The non-trivial solution of this equation is:

$$\alpha = \frac{s \cdot \hat{s}}{\|s\|^2}$$

So we obtain the following definition of SI-SDR:

$$SI\text{-}SDR(s, \hat{s}) = 10 \log_{10} \left( \frac{\|\frac{s \cdot \hat{s}}{\|s\|^2} s\|^2}{\|\frac{s \cdot \hat{s}}{\|s\|^2} s - \hat{s}\|^2} \right)$$

> ***Highlight:*** SI-SDR is simpler than the original SDR metric and can be used as a loss function in ML models. The scale-invariance also prevents the ML model from "taking a shortcut" and applying a time-invariant gain to reduce its loss (as is possible when using SNR as a loss function, for example).
>
> Unlike SDR, however, SI-SDR cannot be adjusted to "allow" distortions other than time-invariant gains. This means that if the ML model introduces some noise or artifacts, they will get penalized even if they do not affect the intelligibility of the output speech signal.

It is worth noting that SI-SDR uses the entire vectors $s$ and $\hat{s}$ instead of breaking them down and comparing them in parts. This may cause a short instantaneous distortion of $\hat{s}$ to skew the whole metric significantly even though it affects intelligibility only in a tiny fraction of the output. This problem has been mitigated in PESQ by "training" the metric parameters so that it closely matches humans' perception of intelligibility. In STOI, the problem has been addressed directly through its mathematical formulation.

## 4.2. Perceptual Evaluation of Speech Quality (PESQ)

PESQ was developed as an ITU standard [16,17] for evaluating the performance of telecommunication devices. It contains a complicated psychoacoustic model that aims to predict how a human subject would evaluate the quality of the degraded (approximated) signal. The calculation of the metric involves the successive application of various filters and transformations to both the original and the degraded signal. At specific points in this transformation process both representations are compared and the differences between them are evaluated. Finally, the evaluations of the "disturbances" are combined in a single linear combination giving the total PESQ score.

The parameters of all evaluations involved in PESQ have been iteratively fine-tuned to best reflect how human subjects perceive the signal. The coefficients of the final linear combination have also been estimated so that PESQ most closely matches the mean opinion score (MOS) of the subjects.

To generate the training data over which the PESQ parameters have been fine-tuned, a set of human subjects has been asked to listen to a list of degraded speech recordings and evaluate them on a scale from 1 (bad) through 5 (excellent). The MOS for each recording has been calculated as the arithmetic average of all subjective scores. PESQ has then been trained to be an estimation of the MOS. Even though the PESQ scores actually range from $-0.5$ to $4.5$, PESQ values below $1.0$ are very rare and occur only if the approximated signal has been greatly distorted [17].

> ***Highlight:*** The calculation of PESQ involves a complicated multi-step process which outputs a number between $-0.5$ and $4.5$. Its parameters have been trained so that its output most closely matches the mean opinion score of people who have been asked to subjectively evaluate the intelligibility of the training data. A Python implementation of PESQ is freely available. PESQ itself is not differentiable, but some efforts have been made to approximate it with an analytical function to be used as a minimization target in ML models [18].

## 4.3. Short-Time Objective Intelligibility

In [19] it is pointed out that choosing the interval of comparison between the clean speech and the approximated signal to be too wide, may cause few regions of high amplitude to dominate the estimated correlation. On the other hand, if the interval is too short, for example 20-30ms, the resolution at low frequencies may be insufficient. Therefore, an analysis window of 333-500ms is suggested as most appropriate.

The STOI metric suggested in [19] is based on correlation coefficients between temporal envelopes of the clean and degraded speech in overlapping segments of length 384 ms. To calculate the correlations, the clean and degraded signals are first decomposed using STFT. The fine-grained STFT coefficients are then grouped together – first along the frequency axis to form one-third octave bins, and then along the temporal axis to form overlapping windows of the specified length. Thus for each signal we obtain one vector representing each one-third octave content in each temporal envelope of length 384 ms.

The vectors representing the degraded signal are then normalized and clipped. Normalization is done in order to handle time-invariant gains applied to the approximated signal. Clipping is justified by the fact that if a portion of the signal is distorted to an extent where it has become unintelligible, any further distortions to this portion will not affect the overall intelligibility of the signal.

The element-wise sample correlation is then calculated between all vectors of the degraded signal and their corresponding vectors from the clean speech signal. Finally, all correlations are averaged to get the total STOI score. As an average of correlation coefficients, the output is a number between $0$ and $1$. The whole process is described in Figure 15.

The authors of [19] claim that STOI is highly correlated with the subjective perceptual intelligibility of the signal. In their tests they demonstrate a correlation of 0.9 between STOI and the intelligibility scores assigned by human subjects.

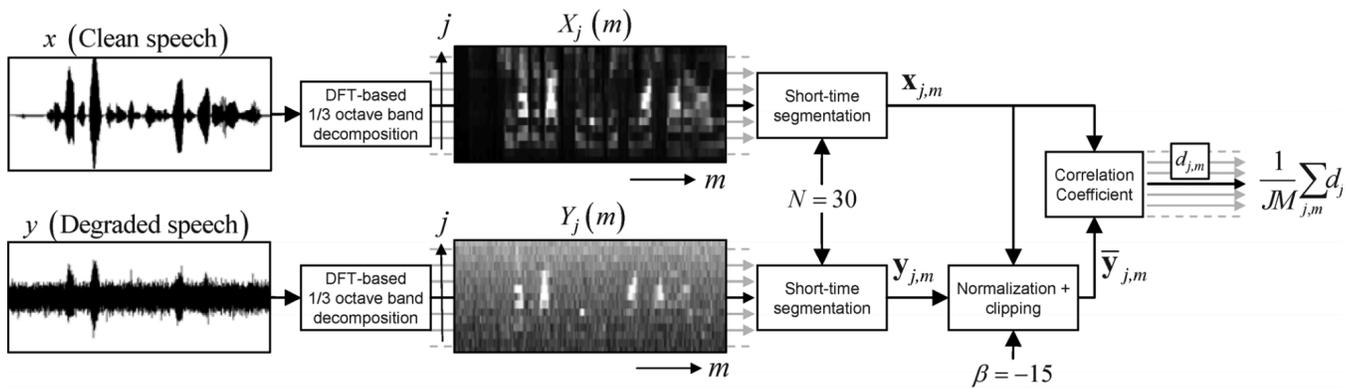

*Figure 15. (Source: [19]) Calculation of STOI. The clean speech $x$ and degraded signal $y$ are first transformed to the frequency domain using a windowed Discrete Fourier Transform (DFT), i.e. STFT, with a Hann window of length 256 samples and 50% overlap. DFT bins are grouped along the frequency axis into ⅓ octave bands. Bins are also grouped along the time axis into temporal envelopes of size $N = 30$ which corresponds to 384ms at 10kHz sampling rate. The degraded signal envelopes are then normalized and clipped using a clipping threshold of $\beta = -15dB$. Finally, the correlation coefficients $d_{j,m}$ between each pair of envelopes is calculated and averaged to obtain the final STOI score.*

*Highlight:* Similarly to PESQ, STOI aims to approximate the subjective evaluation of intelligibility performed by human subjects. Its implementation is simpler than that of PESQ and it is also freely available as a Python package. Separate Python packages exist which contain implementations of STOI as a loss function to be used in ML models.

# 5. The Speaker Isolation Problem

In this section we examine a specific voice processing problem called *speaker isolation*, as well as a class of ML models typically employed for solving it. In this context we propose an experimental design for the comparative evaluation of different signal decomposition methods applied prior to the ML model itself. We also provide experimental results based on an experiment performed according to the proposed design.

## 5.1. Problem Overview

The problem of speaker *recognition* can be divided into three specific tasks [20]:
- Identification: given a voice recording, the task is to determine the identity of the speaker by comparison to a precompiled dataset of known speakers.
- Detection/Verification: given a voice recording and an identity claim, the task is to verify that the speaker is truly the one who they claim to be.
- Segmentation and clustering: given a recording of multiple overlapping voices, the task is to produce one or more output recordings each of which isolates only one of the voices.

The input for all of these tasks is an audio signal (single- or multi-channel) and the output is either a classification result (for identification/verification) or one or more transformed audio signals (for segmentation and clustering tasks).

The specific problem of single-channel speaker *isolation* is a type of segmentation task described as:
- The input signal contains a single audio channel.
- The input signal contains a mixture of speech coming from a (known or unknown) number of different speakers.
- As an additional input, we have a "clean" speech sample (reference) recorded in advance by a given person.
- The task is to transform the input mixture and produce an output signal which contains only the speech of the target person.

This is a variant of what is known as "the cocktail party problem", a term coined by E. Colin Cherry in 1953 [21]. This problem is still considered largely unsolved although there exist machine learning models which provide decent results under additional assumptions [22–24]. In recent years with the development of automated voice assistants and personal hearing aids, the interest in solving the cocktail party problem has been renewed.

## 5.2. Applications

Voice assistants such as Siri (by Apple), Alexa (by Amazon) and Cortana (by Microsoft) are gaining popularity as modern and uniform interfaces to various hardware and software solutions. Their most typical usages include gaining quick access to factual data from the Internet as well as controlling Internet of Things (IoT) devices and systems. The voice recognition capabilities of these assistants have improved during recent years but they still work poorly in noisy environments, especially in public places with multiple background voices. This limits their potential applications, especially in the area of public services. For example, a voice assistant, possibly mounted on a mobile robot, could be deployed as a concierge or waiter at a restaurant or as a tour guide at a museum. Such applications, however, require satisfactory solutions to the cocktail party problem.

Recent advancements in personal hearing aids also rely on speaker clustering and isolation. For example, in [25] a hearing device is proposed, which tracks the brainwaves of the person who is wearing it and based on that determines which voice the person is focusing on. The device then performs clustering of the incoming audio signal in real-time and amplifies the target voice so that the person can hear it better. Such devices can greatly improve the quality of life of hearing-impaired people, but their efficiency is also limited by the accuracy of the available solutions to the cocktail party problem.

## 5.3. Mathematical Formulation

The input for speaker isolation problems is a finite discretely sampled signal of length $L$, $x \in \mathbb{R}^L$ which consists of speech from $n$ speakers, represented by signals $s_i \in \mathbb{R}^L, i = 0, ..., n-1$, mixed together and not available separately:

$$x = \sum_{i=0}^{n-1} s_i$$

The task is to infer one of the signals in the mixture, say $s_0$, given a reference sample $r \in \mathbb{R}^{L_r}$ where $L_r$ is the length of $r$ which is typically smaller than $L$. The reference sample is recorded in advance by the corresponding speaker.

Another variation of the problem requires the inference of *all* signals $s_i, i = 0, ..., n-1$ without the presence of additional reference signals in the input (speaker clustering). The authors of [26], also assume an upper bound on the number of speakers $n$ as an additional simplification to the task.

## 5.4. Solutions Overview

*State-of-the-art source separation methods typically take the route of estimating masking operations in the time-frequency domain (even though there are approaches that operate directly on time-domain signals and use a DNN [deep neural network] to learn a suitable representation from it)* [27].

*Masking* involves element-wise multiplication of the time-frequency[5] (TF) representation of the signal with a matrix called a *mask*. The purpose of the mask is to diminish or eliminate some of the TF components of the signal and preserve others, thus modifying the signal structure. After inverting the masked TF representation of the signal back to the time domain, we obtain a modified version of the input which, if the mask has been chosen correctly, should be the desired solution. Thus, the ML models utilized in this class of algorithms are used to approximate the optimal mask of the solution. The whole process is outlined in Figure 16.

If we have the input and output signals in advance, say from a training dataset, we would be able to calculate the *ideal* mask by taking the TF representations of both signals and performing some operations on them. The type of operations we would perform depends on the type of mask we would like to obtain.

---

[5] The term "time-frequency domain" is typically used to denote the domain of the STFT representation of a signal. The representations generated by wavelet transformations are considered to be in the "time-scale domain". However, for the sake of brevity here we will use the term "time-frequency domain" in a more general sense which also includes the time-scale domain.

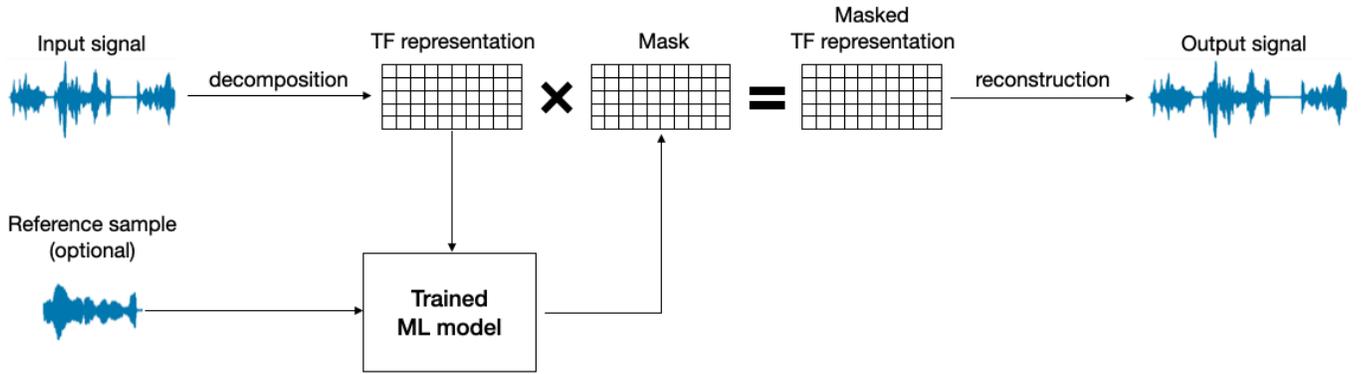

*Figure 16. Typical structure of an algorithm for solving source separation problems. The input signal is decomposed in the time-frequency (TF) domain. Then an ML model is used to generate a mask. After element-wise multiplication of the mask and the TF representation of the signal, the masked TF representation is reconstructed to obtain the output.*

### 5.4.1. Ideal Masks

Let $x \in \mathbb{R}^L$ be a discrete (sampled) signal and let $D$ be an invertible discrete linear transformation, such as STFT. Let us also assume that $x$ consists of some signal of interest $s$ which we would like to isolate (e.g. the voice of a target speaker) and some additive noise $n$, i.e., $x = s + n$. From the linearity of $D$ we obtain:

$$X := D(x) = D(s) + D(n) = S + N$$

For simplicity, let us assume that $D$ produces rectangular matrices, i.e. $X, S, N \in \mathbb{C}^{F \times T}$ for some $F$ and $T$.

As mentioned above, masks are matrices having the same shape as the TF representation of the input signal. Depending on the element type, a mask $M$ can be either binary, i.e., $M \in \{0,1\}^{F \times T}$, or continuous – $M \in \mathbb{R}^{F \times T}$ (or $\mathbb{C}^{F \times T}$ in the more general case). To apply a mask to the signal $X$, we use the element-wise multiplication operator $\circ$. An *ideal* mask $M$ of $s$ is, intuitively, a mask for which the approximation $\hat{s} = D^{-1}(X \circ M)$ approximates $s$ as much as possible. The criteria about whether $\hat{s}$ is a "good" approximation of $s$ may differ depending on the task at hand. Let us now consider two particular examples of ideal masks.

The *ideal binary mask* (IBM) of $s$ is $IBM^s \in \{0,1\}^{F \times T}$ defined as:

$$IBM^s_{f,t} = \begin{cases} 1, & \text{if } |S_{f,t}| - |N_{f,t}| \geq \theta \\ 0, & \text{otherwise} \end{cases}$$

where $\theta$ is a parameter of the mask, typically set to $0$. Put simply, the IBM maps a $1$ to TF bins[6] where the target signal energy dominates and a $0$ to bins where noise dominates.

While there is only one type of binary mask, multiple continuous masks have been developed. Here we will examine one of them - the ideal ratio mask (IRM), only to highlight the differences between IBM and continuous masks.

The *ideal ratio mask* (IRM) of $s$ is $IRM^s \in \mathbb{R}^{F \times T}$ defined as:

$$IRM^s_{f,t} = \frac{|S_{f,t}|^2}{|S_{f,t}|^2 + |N_{f,t}|^2}$$

IRM computes the ratio between the target signal energy and the total energy, thus mapping a number between $0$ and $1$ to each TF bin. IRM and other ideal continuous masks typically lead to more precise reconstruction of the target signal than IBM. However, when a mask is being used as the approximation target of an ML model, there are other characteristics which also need to be considered. They are outlined in the following section.

### 5.4.2. Approximation Accuracy

Clearly, if we do not know $s$ in advance, we cannot calculate its ideal masks, so these masks are typically the target of statistical inference. The question whether a given type of mask is a "good" target for statistical inference does not have an easy answer. Intuitively, IRMs (and other continuous masks in general) work better than IBMs because IBMs either eliminate or preserve each TF bin while IRMs are able to preserve the information in each bin only partially. In other words, even if we have an ML model which constructs $IBM^s$ precisely, this still may not lead to a perfect reconstruction of the target signal $s$.

---

[6] The term *bin* (also *TF bin* or *T-F unit*) is often used to denote the position of a single element of the TF representation of a signal. For example, in the cases where the TF representation is a rectangular matrix, each "bin" is identified by a pair of indices specifying its row and column position respectively.

Even though IBMs do not always lead to perfect reconstruction of the output, it is shown in [28] that they are a reasonable choice of inference targets for source separation ML models. One of the reasons for this is that IBM estimation requires only binary decisions, which makes applicable a host of classification and clustering methods [28]. In ML terms, estimating an IBM can be treated as a classification problem while estimating an IRM (or other real- or complex-valued type of mask) is a regression problem.

Thus single-channel speaker isolation can be transformed from a regression to a classification task where we need to determine the class of each bin (e.g. "noise" or "speech") by inferring the IBM. For the remainder of this work, we will focus on this bin-classification type of algorithms.

Now that we have chosen to work with binary masks, it makes sense to attempt to analyze and quantify the "imperfections" produced by IBMs. In fact, these imperfections can be treated not as flaws of the mask itself, but of the decomposition. If the decomposition $D$ were able to perfectly separate the signal $S$ from the noise $N$, then for each bin $(f, t)$ we would have either $X_{f,t} = S_{f,t}$ or $X_{f,t} = N_{f,t}$. In this case $X \circ IBM^s$ would be exactly equal to $S$ and thus we would obtain a perfect reconstruction $\hat{s} = D^{-1}(X \circ IBM^s) = D^{-1}(S) = s$. So the quality of the IBM is indeed determined by the resolution power of the decomposition $D$, i.e. "how well" is $D$ able to separate the signal from the noise.

In the following section we will perform an experiment which attempts to quantify the resolution power of different signal decomposition methods by performing IBM-based reconstruction and evaluating the result in terms of speech intelligibility.

## 5.5. Experimental Design

The objective of the following experiment is to estimate and compare the resolution power of STFT and Wavelet decompositions with various parameters.

### 5.5.1. Input Data

The experiment uses the Speaker Recognition Audio Dataset [29] which contains multiple recordings from 50 different speakers. Each recording contains pure speech from the corresponding speaker, without additional speech, background music or other significant background noises. Most recordings are precisely one minute long.

Multi-speaker mixtures are generated from this dataset as follows:
1. $n$ distinct speakers are chosen at random. The first of them is designated as the target speaker for the mixture.
2. One recording from each speaker is chosen at random. Each recording is loaded as an array of real-valued samples – $s_0, s_1, \ldots, s_{n-1}$ respectively, $s_0$ being the sample of the target speaker.
3. All sample arrays are padded with zeros at the end so they become of equal length.
4. All samples are summed together to form a mixture:

$$m = \sum_{i=0}^{n-1} s_i$$

### 5.5.2. Experimental Procedure

For each mixture $m$ and each decomposition method $D$:
1. Calculate the decomposition $M = D(m)$ of the mixture $m$.
2. Calculate the decompositions $S^i = D(s_i)$ of each recording $s_i$ from which the mixture $m$ is composed.
3. Calculate the ideal binary mask of the designated target speaker $s_0$ of the mixture:

$$IBM^0_{f,t} = \begin{cases} 1, & \text{if } |S^0_{f,t}| - |\sum_{j>0} S^j_{f,t}| \geq 0 \\ 0, & \text{otherwise} \end{cases}$$

4. Calculate the approximation $\hat{S}^0 := M \circ IBM^0$ of the target signal decomposition using the ideal binary mask.
5. Calculate the approximation of the target signal $\hat{s}_0 = D^{-1}(\hat{S}^0)$ by inverting its approximated decomposition.
6. Estimate the accuracy of the approximation $\hat{s}_0$ compared to $s_0$ by calculating various metrics.

### 5.5.3. Decomposition Parameters

The decompositions compared in this experiment are STFT, WT and WPT. Their parameters were picked using a grid search, i.e., by defining a list of viable values for each parameter and then testing all possible configurations. For STFT the following parameters were varied:
- Window type: Hann or rectangular

- Window size: 5ms, 10ms, 16ms, 25ms, 32ms, 50ms, 100ms, 120ms (these are values found in various research papers and/or used in existing published voice processing algorithms).
- Hop size: 25%, 50%, or 75% of window size.

For WT and WPT the grid search includes the following parameters:
- Wavelet family: all discrete wavelet families supported by version 1.1.1 of the PyWavelets library.
- Vanishing moments: ranging from 1 up to 20 (not all wavelet families support each of these values).
- Levels of decomposition: ranging from 1 up to the maximum $\log_2 L$ where $L$ is the length of the signal.

WPT is performed using full binary tree expansion.

### 5.5.4. Evaluation of Decomposition Configurations

For each configuration, STOI, PESQ, and SI-SDR are calculated and averaged over 10 training examples. The average calculation time for each decomposition is also calculated. Although it can vary greatly depending on the hardware being used, it can be seen as an indication of how fast or slow one decomposition is compared to the others.

## 5.6. Experimental Results

An implementation of the experiment can be found in the following Git repository:

https://github.com/rganchev/speech-signal-processing-for-ml

The following table contains the results of the experiment. Only the three best-performing configurations for each decomposition type are included here, for brevity. The best value in each metric column is in bold font.

| Decomp. | Params | Average metrics | | | |
|---|---|---|---|---|---|
| | | STOI | PESQ | SI-SDR | time (s) |
| STFT | 5ms Hann window 2.5ms hop size | 0.914 | 2.696 | 9.569 | 0.311 |
| | 32ms Hann window 8ms hop size | 0.953 | 2.689 | 13.636 | 0.332 |
| | 50ms Hann window 25ms hop size | **0.957** | 2.689 | **13.800** | 0.166 |
| Wavelet | Symlet 8 vanishing moments | 0.902 | 2.770 | 9.667 | **0.085** |
| | Coiflet 8 vanishing moments | 0.909 | 2.748 | 9.793 | 0.249 |
| | Coiflet 15 vanishing moments | 0.912 | 2.720 | 9.863 | 0.509 |
| Wavelet Packet | Symlet 8 vanishing moments | 0.941 | **2.970** | 12.200 | 0.358 |
| | Coiflet 8 vanishing moments | 0.948 | 2.968 | 12.707 | 1.042 |
| | Coiflet 15 vanishing moments | 0.950 | 2.961 | 12.965 | 2.017 |

## 5.7. Analysis of Experimental Results

The above results were produced using the *ideal* binary mask so they represent an upper limit of the accuracy we could obtain by approximating the IBM with an ML model using the corresponding decomposition method. There are several conclusions we can draw from these results.

Firstly, it is evident that even though STOI and PESQ are both claimed to correlate with subjective MOS, there are variances in their values which can lead to discrepancies. In a practical setting we would probably choose only one of these metrics as the "source of truth" while building the ML model, in order to avoid handling such ambiguities.

Another important observation is that, if fine-tuned correctly, all used decomposition methods can lead to decent results:
- The most stable scores (not very sensitive to configuration changes) seem to be achieved by WPT. Even though WPT is also the slowest transformation on average, there are configurations (such as the Symlet with 8 vanishing moments, for example) which are comparable in speed with the alternative decomposition methods.
- STFT seems to achieve the best overall scores but it also seems to be sensitive to changes in the window parameters. From this experiment alone we cannot conclude whether the best performing STFT configurations are in fact the most suitable ones for use in voice processing tasks in general, because it may be the case that they have been overfitted to work well only with the used dataset. However, as mentioned in previous sections, other research papers in the area also indicate that similar STFT configurations work best when analyzing speech.
- Wavelet Transform is the fastest decomposition on average and it seems to produce higher PESQ scores than STFT, but its STOI and SI-SDR are relatively low. Nevertheless, if speed is of the essence, WT can be a reasonable choice.

From this experiment we can conclude that there are multiple configurations which can work similarly well in this specific context. STFT with a 50ms Hann window and 25ms hop size seems to be the best choice overall, providing highest STOI and SI-SDR scores and having the second fastest average decomposition time. On the other hand, if we choose PESQ to be the "source of truth" metric, then a wavelet decomposition using a Symlet with 8 vanishing moments would be most appropriate. Whether we would use this wavelet via WT or WPT depends on whether speed is more critical than accuracy.

It is worth mentioning again that this experiment compares the decomposition methods in a specific context and the results should not be generalized for other types of voice processing tasks or even for algorithms which do not use ideal binary masks.

# 6. Conclusions and Further Work

In this work we have provided synthesized information which can help ML engineers to:
- Build a better intuition about how STFT, WT and WPT work;
- Make informed predictions about which decomposition method should be more appropriate in a particular setting;
- Compare different speech intelligibility metrics and choose the most appropriate one for a particular task;
- Devise and build experiments which compute quantitative metrics for more precise comparison of the efficiency of signal transformation methods in the context of a specific task;
- Fine-tune the parameters of a chosen signal decomposition method.

All of the analyses considered so far take place *before* the engineer has started building the relevant ML model. Further analysis can be performed if the model is already known. For example, the adjacency considerations described in Figure 9 may not be relevant to the chosen statistical inference method or maybe it can be shown experimentally that the disrupted adjacency of bins does not affect the model's performance significantly. Thus it makes sense to implement the ML model in a way which allows signal transformations to be swapped, evaluated, and compared after the model is already operational.

The described case study of the speaker isolation problem can also be extended by implementing a specific ML model and comparing how it performs when using STFT, WT or WPT respectively. This further analysis can reveal other considerations which may make one set of signal transformation parameters more appropriate than others.

It is also worth mentioning that the provided lists of available signal transformation methods and speech intelligibility metrics is by far not a complete one. For example, constant-Q transforms and mel-spectrograms have been used as extensions of the Fourier transform with the motivation that they better resemble the human auditory system. Indeed, humans can distinguish lower frequencies such as 500-1000Hz but can hardly tell the difference between higher-pitched sounds such as 10,000-10,500Hz. Unlike human perception, STFT has a uniform frequency resolution across the whole frequency spectrum. This makes it reasonable to look for alternatives or extensions of STFT such as the aforementioned constant-Q transforms and mel-spectrograms. The use of these has however become less common in the area of machine learning, since they reduce output quality, and deep learning does not require a compact input representation that they would provide in comparison to the STFT [27].

Regarding speech intelligibility metrics, there is also a wide variety of alternatives. For example PESQ, a.k.a. ITU recommendation P.862, has a successor – ITU-P.863, called Perceptual Objective Listening Quality Analysis (POLQA). The latest version of POLQA was released in 2018. However, to the best of the author's knowledge, its implementation is not freely available. Other metrics also exist, such as the Speech Intelligibility Index (SII), Speech Intelligibility In Bits (SIIB), Hearing-Aid Speech Quality Index (HASQI), etc. These are all different approaches to representing the notion of intelligibility mathematically. Deciding which one is most accurate is not easy and is mostly subjective. PESQ, STOI and SI-SDR may not be the best metrics available but are certainly among the most popular ones at the time of this writing.